\documentclass[journal]{IEEEtran}

\usepackage{amsmath}
\usepackage{algorithm}
\usepackage{algorithmicx}
\usepackage{algpseudocode}
\usepackage{makecell}
\usepackage{amsfonts}
\usepackage{graphicx}
\usepackage{multirow}
\usepackage{booktabs}
\usepackage{color}
\usepackage{url}
\usepackage{hyperref}
\hypersetup{
	colorlinks = true, %Colours links instead of ugly boxes
	urlcolor  = cyan, %Colour for external hyperlinks
	linkcolor = blue, %Colour of internal links
	citecolor = red %Colour of citations
}
\usepackage{soul}
\allowdisplaybreaks[4]

\begin{document}

	\title{Computation-power Coupled Modeling for IDCs and Collaborative Optimization in ADNs}
	%
	%
	% author names and IEEE memberships
	% note positions of commas and nonbreaking spaces ( ~ ) LaTeX will not break
	% a structure at a ~ so this keeps an author's name from being broken across
	% two lines.
	% use \thanks{} to gain access to the first footnote area
	% a separate \thanks must be used for each paragraph as LaTeX2e's \thanks
	% was not built to handle multiple paragraphs
	%
	
	\author{Chuyi~Li,~\IEEEmembership{}
		Kedi~Zheng,~\IEEEmembership{Member,~IEEE,}
		Hongye~Guo,~\IEEEmembership{Member,~IEEE,}
		Chongqing~Kang,~\IEEEmembership{Fellow,~IEEE,}
		and~Qixin~Chen,~\IEEEmembership{Senior~Member,~IEEE}% <-this % stops a space
		\thanks{Manuscript received 13 February 2023; revised 23 July 2023; accepted 26 September 2023. Date of publication 3 October 2023; date of current version 23 April 2024. This work was supported in part by the National Key Research and Development Program of China under Grant 2021YFB2401200, and in part by the National Natural Science Foundation of China under Grant U2066205 and Grant 52107102. Paper no. TSG-00203-2023. \textit{(Corresponding author: Qixin Chen.) }}%
		\thanks{The authors are with the State Key Laboratory of Power Systems, Department of Electrical Engineering, Tsinghua University, Beijing 100084, China (e-mail: qxchen@tsinghua.edu.cn).}% <-this % stops a space
        \thanks{Color versions of one or more figures in this article are available at https://doi.org/10.1109/TSG.2023.3321376. }
        \thanks{Digital Object Identifier 10.1109/TSG.2023.3321376}
	}
	
	% note the % following the last \IEEEmembership and also \thanks - 
	% these prevent an unwanted space from occurring between the last author name
	% and the end of the author line. i.e., if you had this:
	% 
	% \author{....lastname \thanks{...} \thanks{...} }
	%                     ^------------^------------^----Do not want these spaces!
	%
	% a space would be appended to the last name and could cause every name on that
	% line to be shifted left slightly. This is one of those "LaTeX things". For
	% instance, "\textbf{A} \textbf{B}" will typeset as "A B" not "AB". To get
	% "AB" then you have to do: "\textbf{A}\textbf{B}"
	% \thanks is no different in this regard, so shield the last } of each \thanks

% that ends a line with a % and do not let a space in before the next \thanks.
% Spaces after \IEEEmembership other than the last one are OK (and needed) as
% you are supposed to have spaces between the names. For what it is worth,
% this is a minor point as most people would not even notice if the said evil
% space somehow managed to creep in.

% The paper headers
\markboth{IEEE Transactions on Smart Grid, VOL. 15, NO. 3, MAY 2024}%
{Shell \MakeLowercase{\textit{et al.}}: Bare Demo of IEEEtran.cls for IEEE Journals}
% The only time the second header will appear is for the odd numbered pages
% after the title page when using the twoside option.
% 
% *** Note that you probably will NOT want to include the author's ***
% *** name in the headers of peer review papers.                   ***
% You can use \ifCLASSOPTIONpeerreview for conditional compilation here if
% you desire.

% If you want to put a publisher's ID mark on the page you can do it like
% this:
\IEEEpubid{\begin{minipage}{\textwidth}
\centering
\vspace{2em}
\copyright~2024 IEEE. Personal use of this material is permitted. Permission from IEEE must be obtained for all other uses, in any current or future media, \\
including reprinting/republishing this material for advertising or promotional purposes, creating new collective works, for resale or redistribution to servers \\
or lists, or reuse of any copyrighted component of this work in other works.
\end{minipage}}

% Remember, if you use this you must call \IEEEpubidadjcol in the second
% column for its text to clear the IEEEpubid mark.

% use for special paper notices
%\IEEEspecialpapernotice{(Invited Paper)}

% make the title area
\maketitle

% As a general rule, do not put math, special symbols or citations
% in the abstract or keywords.
\begin{abstract}
	The batch and online workload of Internet data centers (IDCs) offer temporal and spatial scheduling flexibility. Given that power generation costs vary over time and location, harnessing the flexibility of IDCs' energy consumption through workload regulation can optimize the power flow within the system. This paper focuses on multi-geographically distributed IDCs managed by an Internet service company (ISC), which are aggregated as a controllable load. The load flexibility resulting from spatial load regulation of online workload is taken into account. A two-step workload scheduling mechanism is adopted, and a computation-power coupling model of ISC is established to facilitate collaborative optimization in active distribution networks (ADNs). To address the model-solving problem based on the assumption of scheduling homogeneity, a model reconstruction method is proposed. An efficient iterative algorithm is designed to solve the reconstructed model. Furthermore, the Nash bargaining solution is employed to coordinate the different optimization objectives of ISC and power system operators, thereby avoiding subjective arbitrariness. Experimental cases based on a 33-node distribution system are designed to verify the effectiveness of the model and algorithm in optimizing ISC's energy consumption and power flow within the system.
\end{abstract}

% Note that keywords are not normally used for peerreview papers.

% For peer review papers, you can put extra information on the cover
% page as needed:
% \ifCLASSOPTIONpeerreview
% \begin{center} \bfseries EDICS Category: 3-BBND \end{center}
% \fi
%
% For peerreview papers, this IEEEtran command inserts a page break and
% creates the second title. It will be ignored for other modes.
\IEEEpeerreviewmaketitle

\section*{Nomenclature}

\addcontentsline{toc}{section}{Nomenclature}

Bold symbols denote vectors or matrices. 

\noindent \textit{Sets and Indices}
\begin{IEEEdescription}
	\item[$ \mathcal{G} $] Set of generators.
	\item[$ \mathcal{T} $] Set of time intervals.
	\item[$ \mathcal{N} $] Set of nodes/vertices.
	\item[$ \mathcal{E} $] Set of branches/edges.
	\item[$ \mathcal{X} $] Set of power system state varibles.
	\item[$ \mathcal{Y} $] Set of ISC workload scheduling varibles.
	\item[$ i,j,k $] General indices.
	\item[$ g $] Index for generators.
	\item[$ n $] Index for IDC.
	\item[$ s $] Index for servers in an IDC.
	\item[$ c $] Index for segments in an IDC.
	\item[$ t $] Index for time intervals.
	\item[$ l $] Index for different types of online tasks.
\end{IEEEdescription}

\noindent \textit{Variables, Parameters and Functions}
\begin{IEEEdescription}
	\item[$ p_j $] Node active power injection at node $ i $.
	\item[$ q_j $] Node reactive power injection at node $ i $.
	\item[$ P_{ij} $] Branch active power flow from node $ i $ to node $ j $.
	\item[$ Q_{ij} $] Branch reactive power flow from node $ i $ to node $ j $.
	\item[$ x_{ij} $] Reactance of branch $ i-j $.
	\item[$ r_{ij} $] Resistance of branch $ i-j $.
	\item[$ V_{i} $] Voltage at node $ i $.
	\item[$ I_{ij} $] Current at branch $ i-j $.
	\item[$ v_{i} $] Squared magnitude of $ V_{i} $.
	\item[$ GP $] Bus-generator incidence status (0 or 1).
	\item[$ IP $] Bus-IDC incidence status (0 or 1).
	\item[$ \ell_{ij} $] Squared magnitude of $ I_{ij} $.
	\item[$ C_{g,t} $] Generation cost of generator $ g $.
	\item[$ C_{0,t} $] Cost of buying electricity from transmission grid.
	\item[$ C_{n,t} $] Cost caused by ISC’s response to DSO. 
	\item[$ B_t $]  Social revenue brought by ISC. 
	\item[$ \varphi_t $] Power factor of ISC.
	\item[$ K_n^C $] Coefficient for IDC's cooling load.
	\item[$ K^{\mathrm{IT}}_{n,s} $] Coefficient for a server's computation load.
	\item[$ U_{n,s} $] CPU utilization rate of a server.
	\item[$ N^\mathrm{IDC} $] The number of IDCs owned by the ISC.
	\item[$ N^{\mathrm{IT}}_n $] The number of servers in IDC $ n $.
	\item[$ C_n $] The number of segments in IDC $ n $.
	\item[$ M_{n,s,t} $] Start-up and shut-down indicator for a server.
	\item[$ \bar M_{n,c,t} $] Start-up and shut-down batch for a segment.
	\item[$ S_{n,s,t} $] On-and-off state indicator for a server.
	\item[$ \bar S_{n,c,t} $] Active server batch for a segment.
	\item[$ N^l $] The VM availability requirement for workload $ l $. 
	\item[$ \Phi_{n,s} $] The number of CPUs of a server.
	\item[$ \phi^l $] The CPU core requirement of a VM for workload $ l $.
	\item[$ a^l_{n,c} $] The VM batch of workload $l$ of a segment
	\item[$ G $] A large enough number.
	\item[$ R^l_{n,s} $] The number of VMs for workload $ l $ on a server.
	\item[$ \lambda^{l}_t $] The overall computation demand for workload $ l $.
	\item[$ \Lambda^{l}_{n,s,t} $] The request flow from a server for workload $ l $.
	\item[$ \eta^l $] Backup rate for computation capability.
	\item[$ \boldsymbol{x} $] Varibles of power system state. 
	\item[$ \boldsymbol{y} $] Varibles of ISC workload scheduling. 
\end{IEEEdescription}

\IEEEpubidadjcol
\noindent \textit{Superscript}
\begin{IEEEdescription}[\IEEEusemathlabelsep\IEEEsetlabelwidth{C, cooling}]
	\item[ISC] Internet Service Company.
	\item[IDC] Internet Data Center.
	\item[C, cooling] Cooling load.
	\item[IT] Denote servers in an IDC. 
	\item[SU] Start-up.
	\item[SD] Shut-down.
	\item[base] Denote base power consumption of an server.
	\item[D] Power demand.
\end{IEEEdescription}

\section{Introduction}
\label{sec:introduction}
\IEEEPARstart{W}{ith} the widespread adoption of the Internet, there has been a significant surge in people's demands for data storage, access, and processing. As of the end of 2019, global network data traffic had already reached a staggering 2.1 zettabytes, marking a twelvefold increase over the span of a decade~\cite{kamiya2020data}. This exponential growth has brought Internet data centers (IDCs) to the forefront. IDCs are facilities housing numerous server clusters, catering to the escalating demands of data-intensive applications and services.

In order to deliver scalable Internet services with minimal latency and high reliability, Internet service companies (ISCs) have deployed a substantial number of geographically distributed IDCs. Within these IDCs, server clusters, and cooling facilities are prominent examples of high-energy-consuming equipment. As of 2016, the global electricity consumption of IDCs had already reached approximately 416 TWh, accounting for 3\% of the world's total electricity consumption. Furthermore, it was projected that this consumption would double every four years~\cite{chen2020internet}. Within China specifically, the overall power consumption of IDCs has been growing at an annual rate of over 10\% for the past decade. By 2030, it is estimated that IDCs' power consumption will exceed 140 TWh, constituting 3.7\% of the total social power consumption. As the proportion of uncontrollable power generation resources, such as photovoltaic and wind power, increases within the power system, the uncertainty on the power generation side becomes more significant. Consequently, the importance of effectively utilizing flexible resources on the user side, specifically within IDCs, becomes more prominent. Incorporating IDCs into demand response projects and harnessing their flexibility has become a crucial area of focus.

The power requirements of IDCs can be categorized into three main components: IT equipment power, cooling system power, and other power. IT equipment power is dedicated to processing network computing tasks assigned by ISC to ensure the quality of service (QoS) for users, while the cooling system provides heat dissipation for the IT equipment. These two aspects of power consumption are closely intertwined, whereas other power consumption, such as lighting, is not directly related to computing services. Typically, the power consumed by IT equipment and the cooling system accounts for around 80\% to 90\% of the total power consumption~\cite{al2008scalable}. The computation demand received by ISCs can be classified into two categories: online workload and batch computing~\cite{kozyrakis2013resource}. Online workload, such as online game services and web searches, is time-sensitive and requires immediate responses from the servers. On the other hand, batch computing, like deep learning tasks, has more flexibility regarding delays and can be scheduled at different times. Techniques such as dynamic voltage and frequency scaling (DVFS)~\cite{grunwald2000policies}, dynamic cluster server configuration (DCSC)~\cite{li2011towards}, and virtual machine (VM) technology~\cite{nathuji2007virtualpower} enable the swift transfer of online workload among IDCs in different geographical locations at a low cost~\cite{tran2015geo}. Additionally, batch computing tasks can be scheduled flexibly in terms of time~\cite{li2016toward}. By employing these techniques, the number and frequency of active servers within an IDC can be dynamically adjusted based on the current demand for computing resources. This highlights the inherent coupling between computation demand and power demand, thereby leveraging the unique demand flexibility of IDCs as power users.

\subsection{Ralated Works}
The modeling and optimization of IDCs' electricity consumption, considering the flexibility of ISC workload scheduling, have been extensively investigated. Real-time pricing (RTP) plays a crucial role in enhancing market efficiency within the operation and planning of power systems. Flexible loads can adjust their power consumption based on the electricity price to achieve energy cost savings. Chen \textit{et al.}~\cite{chen2013electric} developed an ISC scheduling model that responds to locational marginal pricing (LMP) while primarily considering the spatial flexibility of IDCs. They also designed a compensation mechanism to incentivize participation. In~\cite{li2014modeling}, both delay-sensitive online workload and delay-insensitive batch computing were simultaneously considered. They established an ISC scheduling model based on day-ahead predictions to minimize energy costs. Building upon this, an IDC model that incorporates spatial scheduling, temporal shifting, and the thermal inertia of the cooling system has been thoroughly studied~\cite{chen2020internetre}. In~\cite{chen2020internet}, the simultaneous coupling of the three regulation capabilities was considered within a comprehensive model, along with the corresponding scheduling infrastructure. However, in these models, the ISC typically acted as a price-taker without fully considering its interaction with the system operator. This limited their ability to fully utilize the flexibility offered by IDCs. To address this limitation, Wang \textit{et al.}~\cite{wang2015proactive} proposed a pricing mechanism that accounted for the impact of spatial regulation on the power system. The aim was to mitigate electricity price fluctuations caused by the spatial load regulation of IDCs. In~\cite{chen2021incentive}, the IDC model was integrated into optimal power flow (OPF) calculations, and an incentive-compatible demand response framework was designed, considering the impact of IDC flexibility on DC power flow optimization. In this framework, the ISC acted as a fully or partially delegate-controlled load. 

While existing studies have primarily focused on the flexibility of large-scale cross-region IDCs, there is an important aspect of IDC flexibility that has been overlooked. Research indicates that there is a significant number of small and medium-scale data centers distributed within cities, accounting for more than 50\% of the total IDC electricity consumption~\cite{vasques2019review}. For example, Alibaba operates 12 electricity and Internet-independent IDCs in the Beijing region, which are interconnected through low-latency and high-speed private networks to provide Internet services. In most cases, the workload of these IDCs can only be scheduled within their respective regions. In this context, geo-distributed IDCs can be seen as a novel type of distributed energy resources (DERs). They have the potential to contribute to distribution systems with renewable generators by reducing network losses, alleviating network congestion, absorbing renewable generation, and more~\cite{us2019designing}. However, the full utilization of this flexibility in distribution network scenarios remains largely unexplored.

In the context of IDC energy consumption modeling, most current studies simplify workload scheduling scenarios and overlook the inherent heterogeneity of workload. Typically, online workload is described in terms of the number of requests received by the ISC per second~\cite{lu2017bulk,liu2011greening}. However, in actual IDC operations, workload scheduling is more intricate. Workload schedulers leverage VM technology to encapsulate online workload into individual VMs. This approach ensures operational security when handling multiple types of workload on a single server and enhances the mobility of online workload. As a result, the minimum scheduling unit for online services becomes a VM~\cite{liu2018elasticity}. In practice, a workload task flow can only be processed by the corresponding host VM, and the deployment of VMs imposes heterogeneity constraints based on the requirements of different types of workload. Consequently, workload scheduling transforms into a bin-packing problem. However, previous research has often overlooked this heterogeneity and oversimplified the workload scheduling process. This oversight leads to an overestimation of flexibility, making it challenging for ISCs to effectively schedule workload in practical applications. 

In most studies exploring the interaction between an ISC and a system operator, there is a common aim to maximize common interests through mutual agreement. For ISCs, efficient utilization of their computing clusters is a top priority. Well-designed scheduling policies can tightly pack workload, minimizing fragmentation and increasing throughput~\cite{li2019deepjs}. This approach can bring several benefits, such as reducing investment in IDC planning and construction, lowering operational costs, and improving disaster tolerance~\cite{radovanovic2022carbon}. Mathematically, the objective is to maximize overall CPU utilization. On the other hand, when considering IDCs as flexible resources for collaborative optimization scheduling, the system operator's optimization goal is typically to minimize power supply costs by appropriately shifting the IDC load. However, these two objectives can sometimes be conflicting. Certain studies have demonstrated that significant benefits in power grid operation can be achieved by allowing CPU utilization to decrease~\cite{ghatikar2012demand}. In previous research, the adverse impact of the ISC's response was usually expressed as an increase in operating costs~\cite{chen2021incentive}. However, CPU utilization and ISC operating costs are two physical quantities with significantly different value ranges. Selecting an appropriate conversion coefficient between them can be a challenging and subjective task. 

\subsection{Contribution}
This paper addresses the limitations of existing studies by proposing a collaborative optimization model that takes into account the interaction between the distribution system operator (DSO) and the ISC in the day-ahead planning scenario. The model takes into account the flexibility of geographic load balancing for online workload with low delay tolerance. The interaction between the ISC and DSO is specifically considered in the day-ahead power flow optimization process. To adequately capture the scheduling flexibility of the ISC, the model appropriately models the heterogeneity of online workload. Furthermore, recognizing the challenges associated with solving the model in large-scale and complex practical scenarios, the initial model is simplified and reconstructed, and an iterative algorithm is designed to solve the reconstructed model efficiently. In addressing the multi-objective nature involving DSO and ISC, the paper adopts the Nash bargaining solution, ensuring an objective selection of the optimal point to avoid subjective bias. The main contributions of this paper can be summarized as follows: 

\begin{enumerate}
	\item Construction of a power consumption model for IDCs that considers the two-step online workload allocation process involving VM creation and workload allocation. By fully accounting for the heterogeneity of online workload, the model accurately describes the flexibility of the ISC.
	\item Establishment of a computation-power coupled optimization model for the DSO and ISC within the day-ahead distribution network OPF scenario. The model is evaluated through some numerical cases. 
	\item Reconstruction of the initial model to address large scheduling scenarios. An iterative algorithm is devised to efficiently solve the reconstructed model, enhancing computational performance.
	\item Introduction of a multi-objective optimization framework from a game theory perspective, allowing for a trade-off between the conflicting objectives of the DSO and ISC. The framework incorporates the Nash bargaining solution to avoid subjective biases. 
\end{enumerate}

The structure of this paper is organized as follows. Section~\ref{sec:online} provides an analysis of the scheduling mechanism and operational scenarios of ISC online workload. In Section~\ref{sec:problem}, the power consumption model of IDC is established, and the computation-power coupled optimization model is constructed. Section~\ref{sec:model} addresses the computational challenges encountered when solving the model directly for large server clusters. An effective algorithm is designed to overcome this issue. Section~\ref{sec:multi} discusses the conflict between ISC and DSO, and a multi-objective optimization framework based on the bargaining game is designed. In Section~\ref{sec:experiments}, experimental cases are established to evaluate the effectiveness of the optimization model and the multi-objective algorithm. Section~\ref{sec:conclusion} is the conclusion of the paper. 

\section{Online Workload Scheduling Mechanism}
\label{sec:online}
Due to the flexible scheduling options provided by Internet computing workload, ISCs have the ability to modify the timing and location of processing Internet computing tasks. To effectively integrate the operations of the data network and power system, it is crucial to reassess the characteristics of energy consumption, regulation, and computing task scheduling in IDCs from the perspective of computation-power coupling. 

\begin{figure}[!t]
	\centering
	\includegraphics[width=0.30\textwidth]{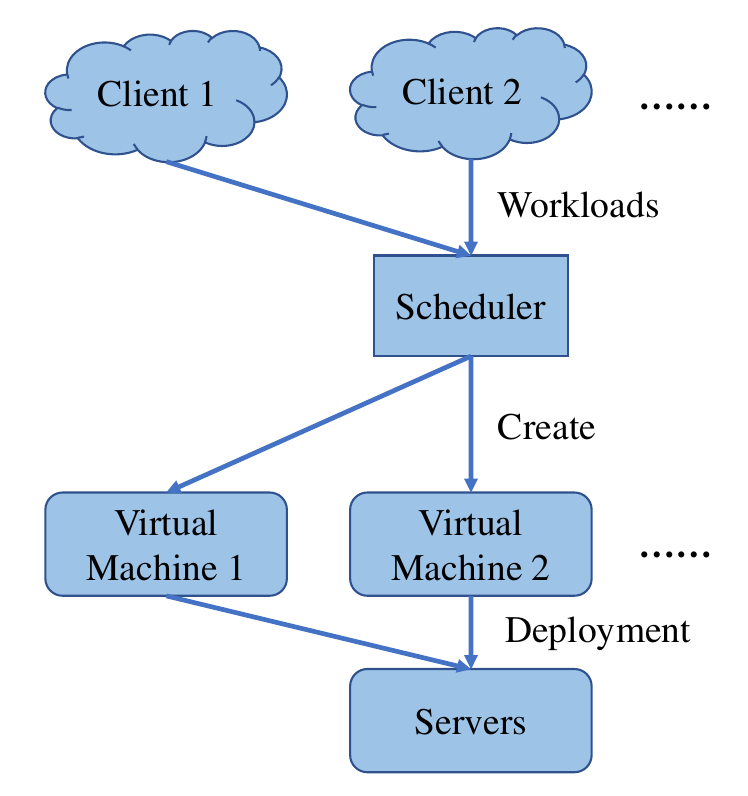}
	\caption{Online workload scheduling mechanism of ISCs.}
	\label{fig:workloadsscheduling}
\end{figure}

This paper primarily focuses on online workload scheduling and its coordination with the distribution system. Online workload typically involves long-running services that require immediate response, and their scheduling generally consists of two steps. Firstly, a group of Virtual Machines (VMs) is created for each client's online workload on physical servers. These VMs can be replicated on a single server or distributed across multiple servers. In practice, clients may specify the requirements for VM sets, such as the number of CPU cores for each VM and the maximum number of VMs that can be replicated on a single physical server. These constraints are referred to as specification requests and availability requests. Secondly, the incoming requests from each client are assigned to the corresponding VMs for processing, as illustrated in Fig.~\ref{fig:workloadsscheduling}. Given the low tolerance for delays in online workload, the waiting time for requests must meet the QoS requirements when workload balancing is achieved. Therefore, the request flow is defined as the number of CPU cores required per unit of time to balance the received workload within the QoS requirements, measured in CPU cores per hour (c/h). The number of CPU cores is an equivalent value, making the request flow a continuous variable. Additionally, a certain margin of CPU resources must be maintained to prevent system crashes, leading to the introduction of VM margin, expressed as a percentage of the maximum CPU utilization. Although the request flow of each client fluctuates from time to time, it constantly occupies a particular scale of VMs under some elastic scaling rules. 

Following the two-step scheduling mechanism, online workload exhibits heterogeneity. Different clients have varying requirements for VM specifications and availability. For clients with computationally intensive tasks, the number of CPU cores in VMs is crucial. Conversely, clients with high disaster recovery requirements prioritize limiting the replication of VMs on a single physical server. Meanwhile, the request flow from each client can only be balanced by its own VMs and is charged accordingly. This leads to workload heterogeneity among different clients.

According to the above analysis, the IDCs' energy consumption model needs to consider the coupling of ISC active and reactive power and online workload traffic in the AC power flow scenario of the distribution networks. At the same time, the heterogeneity should be considered to model the scheduling mechanism, including VM specification, availability, workload request flow, and VM margin. The computation-power coupling model is discussed in the next section. 

\section{Problem Formulation}
\label{sec:problem}
\subsection{Computation-power Coupling of ISC}
Considering that an ISC oversees numerous IDCs linked to grid nodes in various locations, the power composition of the ISC and IDCs is as follows:
\begin{subequations}
	\label{con:1-ISC-power}
	\begin{align}
		& P^{\mathrm{ISC}}_t = \sum_{n=1}^{N^{\mathrm{IDC}}} {P^{\mathrm{IDC}}_{n,t}} \label{con:1-ISC-power-a}  \\
		& Q^{\mathrm{ISC}}_t = P^{\mathrm{ISC}}_{t} \tan{\varphi_{t} } \label{con:1-ISC-power-b} \\
		& P^{\mathrm{IDC}}_{n,t} = P^{\mathrm{IT}}_{n,t} + P^{\mathrm{cooling}}_{n,t} + P^{\mathrm{other}}_{n,t} \label{con:1-ISC-power-c}  \\
		& P^{\mathrm{cooling}}_{n,t} = K^{\mathrm{C}}_n P^{\mathrm{IT}}_{n,t}. \label{con:1-ISC-power-d}
	\end{align}
\end{subequations}
Equation~(\ref{con:1-ISC-power-a}) states that the power of the ISC is the aggregate of all the IDCs' power. Equation~(\ref{con:1-ISC-power-b}) represents the relationship between the active and reactive power consumed by the ISC. It is assumed that each IDC implements measures for reactive compensation, maintaining a fixed power factor of the building denoted as $\varphi_t$. Consequently, the overall power factor of the ISC is also $\varphi_t$. Equation~(\ref{con:1-ISC-power-c}) reveals the division of IDC power into three components: IT equipment power, cooling facility power, and other power. Finally, Equation~(\ref{con:1-ISC-power-d}) utilizes the linear load factor model, proposed by Li \textit{et al.} in~\cite{li2011towards}, to describe the energy consumption of the cooling facilities.

Within each IDC, there is a large number of server clusters (IT equipment). The power composition of the servers is as follows: 
\begin{subequations}
	\label{con:2-IT-power}
	\begin{align}
		& P^{\mathrm{IT}}_{n,t} =  \sum^{N^{\mathrm{IT}}_n}_{s=1}{P^{\mathrm{IT}}_{n,s,t} } \label{con:2-IT-power-a} \\
		& P^{\mathrm{IT}}_{n,s,t} = P^{\mathrm{SU}}_{n,s}M^{\mathrm{SU}}_{n,s,t} + P^{\mathrm{SD}}_{n,s}M^{\mathrm{SD}}_{n,s,t}  \label{con:2-IT-power-b} \\
		& + P^{\mathrm{base}}_{n,s}S_{n,s,t} + K^{\mathrm{IT}}_{n,s}U_{n,s,t}  \notag \\
		& M^{\mathrm{SU}}_{n,s,t},M^{\mathrm{SD}}_{n,s,t},S_{n,s,t} \in \left\{0,1\right\}  \label{con:2-IT-power-c} \\
		& M^{\mathrm{SU}}_{n,s,t} - M^{\mathrm{SD}}_{n,s,t} = S_{n,s,t} - S_{n,s,t-1} \label{con:2-IT-power-d}
	\end{align}
\end{subequations}
Equation~(\ref{con:2-IT-power-a}) indicates that the power of IT equipment in IDCn is the summation of all servers within that IDC. Equation~(\ref{con:2-IT-power-b}) describes the power composition of an individual server. The first two terms represent the average power consumption during server start-up and shut-down. These processes involve fixed electricity consumption and can be converted into average power values over the entire time interval. In cases where optimization intervals are significantly longer than start-up and shut-down processes (e.g., 1-hour optimization interval), this average power value may be small enough to be negligible. However, in certain scenarios, studies have imposed constraints on the number of server on-off transitions~\cite{li2014modeling}. In such cases, these terms can act as penalty terms, penalizing the objective function when servers are frequently started up or shut down to limit the number of on-off transitions. The third term represents the fixed power consumption in the active state, while the last term denotes the dynamic power consumption associated with the server's CPU utilization $U_{n,s,t}$. The linear server energy consumption model proposed in~\cite{li2014modeling} is adopted here. When a server is active, CPU power consumption constitutes the primary dynamic component. Assuming fixed frequency and voltage, this power consumption is approximately proportional to the CPU utilization rate, while the power consumption of other components remains relatively constant~\cite{qureshi2009cutting}. The intercept $P^{\mathrm{base}}_{n,s}$ and slope $K^{\mathrm{IT}}_{n,s}$ can be easily obtained by fitting historical operation data. Equations (\ref{con:2-IT-power-c}) and (\ref{con:2-IT-power-d}) establish the coupling relationship of the state indicators.

The IDC also imposes constraints on server management:
\begin{equation}
	\label{con:3-server-manage}
	0 \leq U_{n,s,t} \leq \bar{U}_{n} , \quad \forall \  t,s
\end{equation}
Equation~(\ref{con:3-server-manage}) restricts the maximum CPU utilization rate of a server to $\bar{U_n}$. This constraint is necessary to prevent issues such as downtime. It is generally recommended to limit the CPU utilization rate to below 90\%~\cite{li2014modeling}.

Based on the modeling described above, the power consumption of IDCs is influenced by the on-off state and CPU utilization of servers, which are, in turn, determined by the workload scheduling mechanism discussed in the previous section. Initially, consider the deployment of VMs, which determines the on-off state of servers. As explained in Section~\ref{sec:online}, the ISC dynamically creates or releases VMs based on the incoming request flow and the corresponding specification and availability requirements. This introduces constraints for VM deployment: 
\begin{subequations}
	\label{con:4-deployment}
	\begin{align}
		& R^l_{n,s,t} \in \mathbb{N} \\
		& R^l_{n,s,t} \leq N^l  \label{con:4-deployment-b} \\
		& \sum^L_{l=1}\left( \phi^l \cdot R^l_{n,s,t} \right) \leq \Phi_{n,s} \label{con:4-deployment-c} \\
		& S_{n,s,t} \leq \sum^L_{l=1}R^l_{n,s,t} \leq G\cdot S_{n,s,t} \label{con:4-deployment-d}
	\end{align}
\end{subequations}
Equation~(\ref{con:4-deployment-b}) indicates that the number of VMs deployed on server $s$ in IDCn for workload $l$ balancing cannot exceed $N^l$ in order to satisfy availability requirements. Equation~(\ref{con:4-deployment-c}) ensures that the sum of CPU cores required by all deployed VMs on server $s$ does not exceed the total number of CPU cores available. Equation~(\ref{con:4-deployment-d}) establishes the relationship between VM deployment and the on-off state of the server. $G$ represents a sufficiently large number. The server is on if any VM is deployed on it, and it is off otherwise.

Then consider the workload assignment to the corresponding VMs, which determines the CPU utilization of servers. As explained in Section~\ref{sec:online}, the request flow from a client can only be assigned to the corresponding VM group. Therefore, the workload balance constraints are imposed:
\begin{subequations}
	\label{con:5-workload}
	\begin{align}
		& \Lambda^l_{n,s,t} \geq 0 \\
		& \sum^{N^\mathrm{IDC}}_{n=1}{\sum^{N^{\mathrm{IT}}_n }_{s=1}{\Lambda^l_{n,s,t}}} = \lambda^l_t  \label{con:5-workload-b}  \\
		& \phi^l \cdot R^l_{n,s,t} \cdot \eta^l \geq \Lambda^l_{n,s,t} \label{con:5-workload-c} \\
		& U_{n,s,t} = \frac{\sum^L_{l=1}{\Lambda^l_{n,s,t}}}{\Phi_{n,s}}  \label{con:5-workload-d} 
	\end{align}
\end{subequations}
The symbol $\Lambda_{n,s,t}^l$ represents the request flow of workload $l$. The constraint in Equation~(\ref{con:5-workload-b}) ensures the balanced distribution of the request flow for workload $l$. Equation~(\ref{con:5-workload-c}) ensures that the workload assigned to server $s$ in IDCn are balanced by the corresponding VMs, with $\eta^l$ denoting the redundancy. Equation~(\ref{con:5-workload-d}) describes the relationship between the allocation of request flow and the CPU utilization rate.

\begin{figure}[!t]
	\centering
	\includegraphics[width=0.45\textwidth]{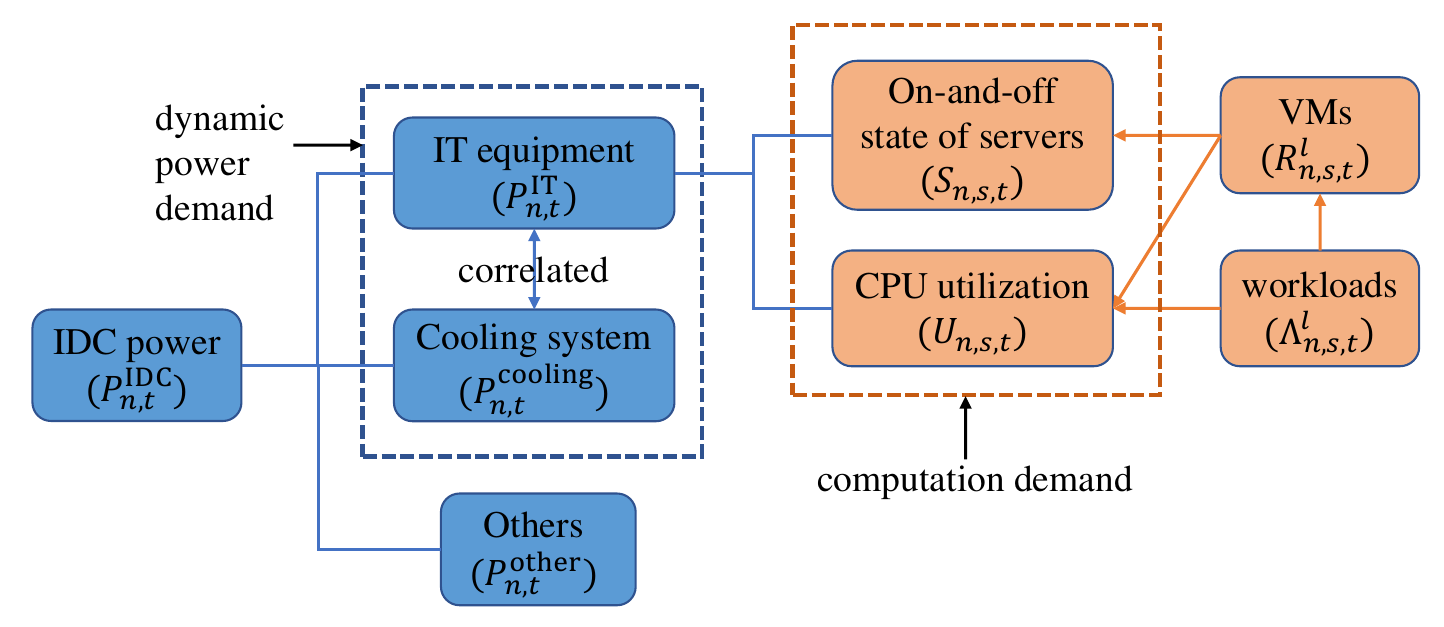}
	\caption{Computation-power coupling relationship of IDC and workload.}
	\label{fig:coupling}
\end{figure}

Equations (\ref{con:1-ISC-power})-(\ref{con:5-workload}) establish the coupling relationship between ISC power consumption and online workload while integrating the scheduling mechanism for ISC VM deployment and workload balance into a single model. This model accurately captures the load flexibility of ISC resulting from the geographic balance of online workload. The logical relationship of each group of constraints is depicted in Fig.~\ref{fig:coupling}. 

\subsection{ISC-DSO Collaborative Optimization Model}
To construct the collaborative optimization model for ISC and DSO, it is necessary to consider the network power flow constraint. For a typical radial network with the network structure $\left(\mathcal{N},\mathcal{E}\right)$, the second-order cone relaxation AC power flow model is adopted~\cite{farivar2013branch}:
\begin{subequations}
	\label{con:6-dist-flow}
	\begin{align}
		&	p_j = \sum_{k:j \rightarrow k}P_{jk} - \sum_{i:i \rightarrow j}\left(P_{ij} - r_{ij}\ell_{ij}\right),   \quad \forall j\in \mathcal{N}  \\
		&	q_j = \sum_{k:j \rightarrow k}Q_{jk} - \sum_{i:i \rightarrow j}\left(Q_{ij} - x_{ij}\ell_{ij}\right), \quad \forall j\in \mathcal{N}   \\
		&	v_j = v_i - 2\left( r_{ij}P_{ij} + x_{ij}Q_{ij}\right) + \left( r_{ij}^2 + x_{ij}^2\right)\ell_{ij}, \\
		&  \qquad \forall \left(i,j\right)\in \mathcal{E}  \notag \\
		&	\underline{p_i} \leq p_i \leq \overline{p_i}, \quad \forall i\in \mathcal{N}   \\
		&	\underline{q_i} \leq q_i \leq \overline{q_i}, \quad \forall i\in \mathcal{N}  \\
		&	\underline{v_i} \leq v_{i} \leq \overline{v_i}, \quad  \forall i\in\ \mathcal{N}   \\
		&	\ell_{ij} \leq \overline{\ell_{ij}}, \qquad \forall \left(i,j\right)\in \mathcal{E}  \\
		&	\left\|{\begin{matrix}
				2P_{ij} \\
				2Q_{ij} \\
				\ell_{ij}-v_i
		\end{matrix}}\right\|_2 \leq
		\ell_{ij}+v_i, \quad \forall \left(i,j\right)\in \mathcal{E}
	\end{align}
\end{subequations}
$j=0$ corresponds to the root node, which is connected to the superior transmission network. At each node, the injection power is determined by the algebraic sum of several components: the load power at that node, the injected power from a connected generator (if any), and the power consumed by the IDC (if any). It should be noted that a node can be connected to at most one generator and one IDC. The injection power at a node can be represented as follows:
\begin{subequations}
	\label{con:7-power-coupling}
	\begin{align}
		&	p_i=GP \cdot p_g-P_i^\mathrm{D}-IP \cdot P_n^\mathrm{IDC}  \\
		&	q_i=GP \cdot q_g-Q_i^\mathrm{D}-IP \cdot Q_n^\mathrm{IDC}
	\end{align}
\end{subequations}
The time subscript $t$ is omitted here. 

Equations (\ref{con:1-ISC-power})-(\ref{con:7-power-coupling}) illustrate the interdependence between the computational demand and power demand of the collaborative ISC and DSO systems when the ISC functions as a flexible load within the distribution system operation. The model takes into account the two-step workload balancing process specific to the distribution system scenario. During the corresponding OPF calculation, the objective function of the problem is typically selected to maximize the total social revenue.
\begin{align}
	\label{obj:initial}
	z_1 = &\sum_{g\in\mathcal{G}, t\in \mathcal{T} } C_{g,t} ( p_{g,t} ) + \sum_{t\in\mathcal{T}} C_0(p_{0,t}) \\
	&+ \sum_{n,t \in \mathcal{T}} C_{n,t}\left(P_{n,t}^\mathrm{IDC}\right) - 
	\sum_{t\in\mathcal{T}} B_t \notag
\end{align}
The first term represents the total generation cost of the DSO. The second term denotes the electricity purchase cost of the DSO from the upper power grid. The third term represents the additional cost caused by the ISC's response to the DSO. Finally, the last term refers to the social revenue brought by ISC's electricity consumption. The ISC-DSO collaborative optimization model is then established as follows:
\begin{align}
	\label{model:initial}
	&\min_{ \begin{matrix}
			\boldsymbol{p}_{\mathcal{G,T}}, \boldsymbol{q}_{\mathcal{G,T}}, \boldsymbol{p}_{0,\mathcal{T}}, \boldsymbol{q}_{0,\mathcal{T}} \\
			R_{n,s,t}, \Lambda^l_{n,s,t}, U_{n,s,t}
	\end{matrix} } z_1 \\
	\text{s.t.: } &(\ref{con:1-ISC-power})-(\ref{con:7-power-coupling}) \notag
\end{align}

\section{Model Reconstruction and Solution Method}
\label{sec:model}
As explained in Section~\ref{sec:problem}, an initial model, Equation~(\ref{model:initial}), is developed to optimize the allocation of computing resources and power flow coordination. However, due to the introduction of AC-OPF constraints in Equation~(\ref{con:6-dist-flow}) and the use of integer variables in Equation~(\ref{con:4-deployment}), the model becomes a mixed-integer second-order cone programming (MISOCP) problem. In practice, the scale of server clusters in IDCs can be enormous, with potentially hundreds of servers in small and medium IDCs, as considered in our paper. The introduction of integer variables for each server to indicate its state significantly increases the size of the model, leading to potential numerical issues. Furthermore, each client may have different CPU core requirements and availability requests, which introduces a new set of integer variables for VM deployment. Consequently, for practical applications, it is imperative to address the computational challenges associated with the model. A model reconstruction method is implemented, and an algorithm is designed to solve the reconstructed model iteratively. 

\subsection{Model Reconstruction}
\label{sec:model-reconstruct}
Based on the previous analysis, the optimization problem becomes challenging due to the numerous possibilities of VM deployment resulting from diverse workload types and unique deployment choices for each server. To address this issue, the concept of homogeneity for both servers and workload is introduced. The homogeneity of servers assumes that the server specifications, such as CPU cores and power efficiency, are not significantly different. In practice, many servers are identical even when located in different IDCs. Therefore, the servers within an IDC can be divided into segments where the server parameters are identical. Furthermore, the VM deployment schemes within a cluster segment exhibit high homogeneity because many types of workload require large batches, and the VM deployment schemes are repeated across servers. As a result, it is unnecessary to introduce separate variables for each server. Additionally, the homogeneity of workload suggests that only a small number of VM deployment schemes are efficient for the optimization objective. For example, when deploying a 2-core VM (VM1) and a 4-core VM (VM2) on an 8-core server, schemes such as 1 VM1 and 1 VM2 are obviously inefficient. By pruning these inefficient schemes, we can simplify the solution space, leading to improved computational efficiency.

A model reconstruction is performed to fully leverage the homogeneity of servers and workload. To eliminate inefficient schemes, an enumeration process is incorporated to replace the VM deployment optimization. Initially, an enumeration of all possible VM deployment schemes is conducted, and inefficient schemes are excluded from further consideration in the optimization process. Each deployment scheme is associated with an integer variable, $\bar S_{n,c,t}$, representing the repeated batch of the scheme on servers. In this context, the servers within each IDC are divided into distinct segments, where machine parameters and VM deployment are identical among the servers. The variable $\bar S_{n,c,t}$ denotes the number of active servers in the segment cluster $c$ of IDC$n$ during time interval $t$. As a result, the VM deployment constraint, Equation~(\ref{con:4-deployment}), is pruned from the initial model~(\ref{model:initial}). Additionally, Equations~(\ref{con:2-IT-power}), (\ref{con:3-server-manage}), and (\ref{con:5-workload}) are revised as follows. 

Assuming that workload is evenly distributed among VMs within the same segment, the CPU utilization of servers in that segment is uniform and denoted as $U_{n,c,t}$. The power composition of the IDC and the relationships among state indicators are revised and presented in batches:
\begin{subequations}
	\label{con:10-IT-power-rev}
	\begin{align}
		& P^{\mathrm{IT}}_{n,t} = \sum^{C_n}_{c=1}{P^{\mathrm{IT}}_{n,c,t} }  \\
		& P^{\mathrm{IT}}_{n,c,t} = P^{\mathrm{SU}}_{n,c}\bar M^{\mathrm{SU}}_{n,c,t} + P^{\mathrm{SD}}_{n,c}\bar M^{\mathrm{SD}}_{n,c,t} \\
		& + P^{\mathrm{base}}_{n,c} \bar S_{n,c,t} + K^{\mathrm{IT}}_{n,c} U_{n,c,t}\bar S_{n,c,t}  \notag \\
		& \bar M^{\mathrm{SU}}_{n,c,t},\bar M^{\mathrm{SD}}_{n,c,t},\bar S_{n,c,t} \in \mathbb{N} \\
		& \bar M^{\mathrm{SU}}_{n,c,t} - \bar M^{\mathrm{SD}}_{n,c,t} = \bar S_{n,c,t} - \bar S_{n,c,t-1}
	\end{align}
\end{subequations}

The server management constraint is thus revised as: 
\begin{equation}
	\label{con:11-server-manage-rev}
	0 \leq U_{n,c,t}\bar S_{n,c,t} \leq \bar{U}_{n}\bar S_{n,c,t} , \quad \forall \  t,c
\end{equation}

The workload balance constraints, considering $a^l_{n,c}$ as the VM batch of workload $l$ allocated to segment $c$ in IDC$n$, are revised as follows:
\begin{subequations}
	\label{con:12-workload-rev}
	\begin{align}
		& \Lambda^l_{n,c,t} \geq 0 \\
		& \sum^{N^\mathrm{IDC}}_{n=1}{\sum^{C_n }_{c=1}{\Lambda^l_{n,c,t}}} = \lambda^l_t  \\
		& a^l_{n,c} \cdot \phi^l \cdot \bar S_{n,c,t} \cdot \eta^l \geq \Lambda^l_{n,c,t}  \\
		& U_{n,c,t}\bar S_{n,c,t} = \frac{\sum^L_{l=1}{\Lambda^l_{n,c,t}}}{\Phi_{n,c}} 
	\end{align}
\end{subequations}

Thus, the reconstructed model is represented as follows: 
\begin{align}
	\label{model:reconstructed}
	&\min_{ \begin{matrix}
			\boldsymbol{p}_{\mathcal{G,T}}, \boldsymbol{q}_{\mathcal{G,T}}, \boldsymbol{p}_{0,\mathcal{T}}, \boldsymbol{q}_{0,\mathcal{T}} \\
			\bar S_{n,c,t}, \Lambda^l_{n,c,t}, U_{n,c,t}
	\end{matrix} } z_1 \\
	\text{s.t.: } &(\ref{con:1-ISC-power}),(\ref{con:6-dist-flow}),(\ref{con:7-power-coupling}),(\ref{con:10-IT-power-rev}),(\ref{con:11-server-manage-rev}),(\ref{con:12-workload-rev}) \notag
\end{align}

Instead of assigning a set of integer variables for each server in Model~(\ref{model:initial}), Model~(\ref{model:reconstructed}) divides the server cluster within each IDC into multiple segments, assuming homogeneity within each segment. In this approach, an integer variable is allocated to determine the batch of the segment. By considering the perspective of segments, the problem scale is significantly reduced, enabling efficient solving of the model.

\subsection{Algorithm Design for Model Solution}
\label{sec:algorithm}
The reconstructed model effectively optimizes the collaboration between ISC and DSO in large server clusters. However, enumerating all possible VM deployment schemes before pruning can lead to numerical challenges. To overcome this issue, an algorithm based on Benders decomposition and column generation is proposed. 

Since Model~(\ref{model:reconstructed}) represents a MISOCP problem, solving it becomes challenging when dealing with large-scale integers. A common approach to address this issue is to separate the complex variables (integer variables) from the simple variables (continuous variables). Intuitively, the variables in the model can be divided into two parts based on their physical interpretation: power system state variables and ISC workload scheduling variables. The former corresponds to continuous variables subject to second-order cone constraints, while the latter corresponds to integer variables subject to linear constraints. These two parts are connected through Constraint~(\ref{con:7-power-coupling}) via the variable $P_{n,t}^\mathrm{IDC}$. Additionally, the objective function $z_1$ can also be split into two parts: $f\left(\boldsymbol{x},P_{n,t}^\mathrm{IDC}\right)$ and $g(\boldsymbol{y})$. As a result, the reconstructed model can be decomposed into a master problem:
\begin{align}
	\label{model:MP0}
	&\min g(\boldsymbol{y}) \\
	\text{s.t.: } &(\ref{con:1-ISC-power}),(\ref{con:10-IT-power-rev}),(\ref{con:11-server-manage-rev}),(\ref{con:12-workload-rev}) \notag
\end{align}
and a subproblem: 
\begin{align}
	\label{model:SP}
	&\min f\left(\boldsymbol{x},\hat P_{n,t}^\mathrm{IDC}\right) \\
	\text{s.t.: } &(\ref{con:6-dist-flow}),(\ref{con:7-power-coupling}) \notag
\end{align}

The master problem is formulated as a mixed-integer linear programming (MILP) problem, while the subproblem is a second-order cone programming (SOCP) problem, both of which can be efficiently solved using commercial solvers. According to Benders decomposition~\cite{geoffrion1972generalized}, the optimal solution of the original MISOCP problem can be approached by iteratively solving the master problem and the subproblem. The approach begins by solving Model~(\ref{model:MP0}) to obtain an initial solution. This initial solution represents the optimal operation of ISC, considering only its own interests. The corresponding power consumption of ISC is denoted as $\hat P_{n,t}^\mathrm{IDC}$. Then, this power consumption is treated as a fixed demand in the subproblem~(\ref{model:SP}), which involves an OPF calculation. In general cases, the subproblem is feasible. An optimality cut can be formulated based on the Benders decomposition method and added to the master problem: 
\begin{align}
	\label{model:MP}
	&\min g(\boldsymbol{y}) + q \\
	\text{s.t.: } &(\ref{con:1-ISC-power}),(\ref{con:10-IT-power-rev}),(\ref{con:11-server-manage-rev}),(\ref{con:12-workload-rev}) \notag \\
	&\text{Benders optimality cuts} \notag \\
	&q \geq 0 \notag
\end{align}
The objective of the DSO is thus partially addressed in the workload scheduling optimization process through a set of linear constraints, as represented in~(\ref{model:MP}). Subsequently, the value of $\hat P_{n,t}^\mathrm{IDC}$ can be updated by solving the updated master problem. This iterative process continues until convergence is achieved.

In practical application, only a few selected deployment schemes are actually involved in determining the final optimal solution. It is unnecessary to search for the optimal point among all possible enumerations. To address this, column generation, a well-known algorithm, is utilized to handle the large-scale MILP problem described earlier. It is worth noting that the reconstructed model has been formulated to directly incorporate column generation. As a result, the enumeration step is bypassed, and the column generation algorithm is directly applied to solve the master problem~(\ref{model:MP}) in each iteration. The algorithm for solving Model~(\ref{model:reconstructed}) is summarized in \textbf{Algorithm~\ref{algo:benders}}.

\begin{algorithm}[ht]
	\caption{Solution method for the reconstructed model}
	\label{algo:benders}
	\begin{algorithmic}[1]
		\State Use branch and price algorithm to solve ISC optimal scheduling problem~(\ref{model:MP0}) and get $\hat P_{n,t}^\mathrm{IDC}$, $\hat{\boldsymbol{y}}$; 
		\State \textit{Initialization} : $\epsilon \leftarrow$ a positive small gap (for example, 0.001), $\mathrm{LB} \leftarrow -\infty$, $\mathrm{UB} \leftarrow +\infty$; 
		\While{$\mathrm{UB} - \mathrm{LB} \geq \epsilon$} \\
		\quad \textbf{(Solve subproblem)}
		\State Solve Model~(\ref{model:SP}) and get $ \hat{\boldsymbol{x}}$; 
		\State Add optimality cuts to master problem; 
		\State $\mathrm{UB} \leftarrow \min \left\{\mathrm{UB}, f\left(\hat{\boldsymbol{x}},\hat P_{n,t}^\mathrm{IDC}\right) + g(\hat{\boldsymbol{y}})\right\}$ \\
		\quad \textbf{(Solve master problem)}; 
		\State Use branch and price algorithm to solve Model~(\ref{model:MP}) and get $\hat q$, $\hat{\boldsymbol{y}}'$, $\hat P_{n,t}^{\mathrm{IDC} '}$;  
		\State $\hat{\boldsymbol{y}} \leftarrow \hat{\boldsymbol{y}}'$; $\hat P_{n,t}^\mathrm{IDC} \leftarrow \hat P_{n,t}^{\mathrm{IDC} '}$; 
		\State $\mathrm{LB} \leftarrow g(\hat{\boldsymbol{y}}) + \hat q$; 
		\EndWhile
		\State Get the optimal point $\left\{\hat{\boldsymbol{x}}, \hat P_{n,t}^\mathrm{IDC}, \hat{\boldsymbol{y}}\right\}$. 
	\end{algorithmic}
\end{algorithm}

Regarding the convergence aspect, the algorithm is structured as a combination of column generation and Benders decomposition, which results in nested iterations. The inner iteration, which corresponds to the master problem, inherits the convergence properties of column generation. On the other hand, the outer iteration follows the convergence characteristics of Benders decomposition. In practical applications, both iterations tend to converge rapidly, ensuring efficient convergence of the algorithm as a whole.

\section{Multi-objective Optimization Framework}
\label{sec:multi}
Equation~(\ref{obj:initial}) presents the objective function of the proposed model, which aims to maximize the total social welfare. In the considered scenario, the response of ISC to DSO is limited to load shifting rather than load reduction. Additionally, the request flow of all online workload must be balanced while ensuring QoS. Consequently, the social revenue generated by ISC electricity consumption remains constant, making the last term in Equation~(\ref{obj:initial}) a fixed value. The first two terms in the objective function represent the cost incurred by the DSO in supplying power, which is dependent on the amount of electricity generated and purchased. The third term accounts for the cost associated with the impact on ISC's interests caused by complying with DSO's control, which can be challenging to evaluate. When an ISC operates independently, the scheduler's optimization objective typically revolves around maximizing CPU utilization on the active server. However, ISC's participation in the collaborative optimization of network power flow affects workload scheduling, potentially conflicting with the goal of maximizing CPU utilization. To further explore the optimization objective, a two-objective optimization problem can be formulated. This highlights the inherent conflict between the optimization objectives of the DSO and the ISC. 

In most occasions, the energy consumption scale of an ISC is limited compared to the whole distribution system load, which makes the objective of DSO seem to be more important. However, fully incorporating the ISC into regulation, regardless of the cost of the ISC being regulated, would be detrimental to the ISC's operations. The benefits gained from the ISC's free electricity usage often outweigh the costs of electricity production itself. On the other hand, solely focusing on the ISC's objective is also inappropriate since, within a certain range, the ISC can dispatch workload without incurring additional costs~\cite{ghatikar2012demand}. Neglecting the DSO's objective would lead to a non-Pareto-efficient solution. Therefore, it is crucial to investigate methods for balancing the objectives of both parties.

$z_2$ and $z_3$ are used to represent their objective functions as follows:
\begin{equation}
	\label{obj:DSO}
	z_2 = \sum_{g\in\mathcal{G}, t\in \mathcal{T} } C_{g,t} \cdot p_{g,t}  + \sum_{t\in\mathcal{T}} C_0 \cdot p_{0,t} 
\end{equation}
\begin{equation}
	\label{obj:ISC}
	z_3 = -\frac{1}{T} \sum^T_{t=1}{\frac{\sum_{l} \lambda^l_t}{\sum^{N^{\mathrm{IDC}}}_{n=1}{\sum^{C_n}_{c=1}{\Phi_{n,c} \cdot \bar S_{n,c,t}}}}}
\end{equation}
Equation~(\ref{obj:DSO}) represents the power supply cost of the DSO, which is modeled using a linear cost function. On the other hand, Equation~(\ref{obj:ISC}) represents the negative average CPU utilization rate of the active server in the ISC. If the request flows $\lambda_t^l$ cannot be reduced for the ISC, the objective functionn~(\ref{obj:ISC}) can be equivalently expressed as the following linear function:
\begin{equation}
	\label{obj:ISC-linear}
	z_4 = \sum^T_{t=1}{\sum^{N^{\mathrm{IDC}}}_{n=1}{\sum^{C_n}_{c=1}{\Phi_{n,c} \cdot \bar S_{n,c,t}}}}  
\end{equation}
The objective function represents the total number of CPU cores activated by ISC.

Thus, the ISC-DSO collaborative multi-objective optimization model can be obtained: 
\begin{align}
	\label{model:multi-obj}
	&\min_{ \begin{matrix}
			\boldsymbol{p}_{\mathcal{G,T}}, \boldsymbol{q}_{\mathcal{G,T}}, \boldsymbol{p}_{0,\mathcal{T}}, \boldsymbol{q}_{0,\mathcal{T}} \\
			\bar S_{n,c,t}, \Lambda^l_{n,c,t}, U_{n,c,t}
	\end{matrix} } \left\{ z_2;z_4\right\} \\
	\text{s.t.: } &(\ref{con:1-ISC-power}),(\ref{con:6-dist-flow}),(\ref{con:7-power-coupling}),(\ref{con:10-IT-power-rev}),(\ref{con:11-server-manage-rev}),(\ref{con:12-workload-rev}) \notag
\end{align}

Generally, when dealing with a multi-objective problem described as Equation~(\ref{model:multi-obj}), a weighting coefficient is commonly used to combine the objectives into a single optimization problem. However, in the context analyzed earlier, the benefits of optimizing CPU utilization are diverse and difficult to quantify, making the selection of a weighting coefficient challenging and subjective. To address this, the bargaining game method is proposed to obtain the optimal solution for the model. In this method, each goal in the multi-objective problem is treated as a competing negotiation unit. These units strive to achieve the best outcome for their respective goals, avoid unfavorable strategies, and ultimately reach a compromise that is acceptable to all negotiation units. In the problem described by Equation~(\ref{model:multi-obj}), two virtual players, the DSO and ISC, is considered. Both players have the same strategy sets, which are the feasible regions of variables $\boldsymbol{p}_{\mathcal{G,T}}$, $\boldsymbol{q}_{\mathcal{G,T}}$, $\boldsymbol{p}_{0,\mathcal{T}}$, $\boldsymbol{q}_{0,\mathcal{T}}$, and $\Lambda^l_{n,c,t}$ determined by Constraints~(\ref{con:1-ISC-power}), (\ref{con:6-dist-flow}), (\ref{con:7-power-coupling}), (\ref{con:10-IT-power-rev}), (\ref{con:11-server-manage-rev}), and (\ref{con:12-workload-rev}). The payoff functions of the two players correspond to the optimization objectives $z_2$ and $z_4$. To determine the equilibrium when the negotiation is completed, the Nash bargaining solution is used, which can be obtained by solving the optimization problem~(\ref{model:nash}): 
\begin{equation}
	\label{model:nash}
	\max_{\left(z_2,z_4\right) \in \mathcal{P}} \left(z_{2,\max }-z_2\right)\left(z_{4,\max }-z_4\right)
\end{equation}
$z_{2,\max}$ and $z_{4,\max}$ represent the maximum possible payoffs of the two players, while $\mathcal{P}$ denotes the Pareto front of the multi-objective problem. The Nash bargaining solution possesses certain key properties, namely Pareto-efficiency, symmetry, and invariance~\cite{Dubey1986Inefficiency}. Pareto-efficiency ensures that the optimal point is efficient in terms of achieving the objectives. Symmetry guarantees fairness to both parties involved. Invariance guarantees the selection of a reasonable optimal solution, even when the physical dimensions and orders of magnitude of the objectives differ significantly. These properties form the basis for selecting the Nash bargaining solution as the optimal point of the model, devoid of subjective arbitrariness.

The estimation of the Pareto front is a crucial step in obtaining the Nash bargaining solution. In Section~\ref{sec:model}, a model solution method was proposed, which involves generating a series of intermediate points during the iteration process. From a multi-objective optimization perspective, \textbf{Algorithm~\ref{algo:benders}} initially identifies the optimal point that represents ISC's primary objective. Subsequently, as optimality cuts are incorporated, the solution gradually shifts towards a specific point on the Pareto front. The convergence point is determined by the weighting coefficient assigned to the two objectives. To obtain a numerical estimation of the Pareto front $\mathcal{P}$, it is necessary to modify the weighting coefficient and repeat the execution of \textbf{Algorithm~\ref{algo:benders}}. This iterative process generates a set of points, and their convex hull provides the numerical approximation of the Pareto front $\mathcal{P}$.

\begin{figure}[!t]
	\centering
	\includegraphics[width=0.45\textwidth]{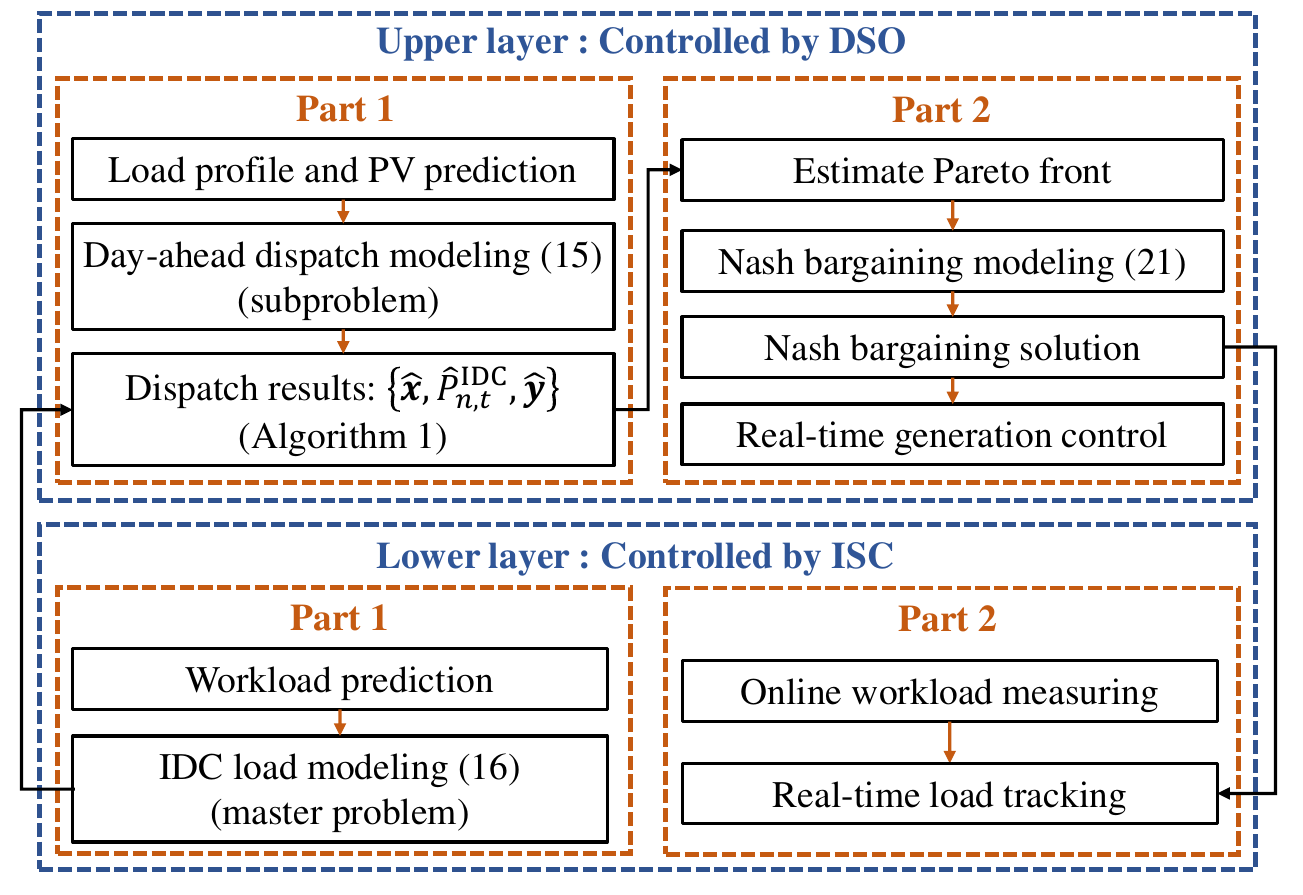}
	\caption{Diagram of ISC-DSO collaborative optimization framework.}
	\label{fig:framework}
\end{figure}

Fig.~\ref{fig:framework} integrates the proposed model and algorithm into the day-ahead optimal dispatch of DSO. The framework consists of two layers: the upper layer controlled by the DSO and the lower layer controlled by the ISC. Initially, both the DSO and ISC forecast the data for the next day, including load, PV output, and workload. These forecasts are utilized to establish the subproblem model and the master problem model, as described in the previous section. DSO summarizes both models and solves them iteratively using \textbf{Algorithm~\ref{algo:benders}}. The obtained results are then forwarded to the second part of the upper layer to estimate the Pareto front. Subsequently, the Nash bargaining model is applied to determine the Nash bargaining solution, representing the optimal operation point. The resulting day-ahead dispatch schedule is subsequently communicated to the generation side and ISC for intra-day real-time tracking.

\section{Case Study}
\label{sec:experiments}
\subsection{Experimental Settings}
\begin{figure}[!t]
	\centering
	\includegraphics[width=0.45\textwidth]{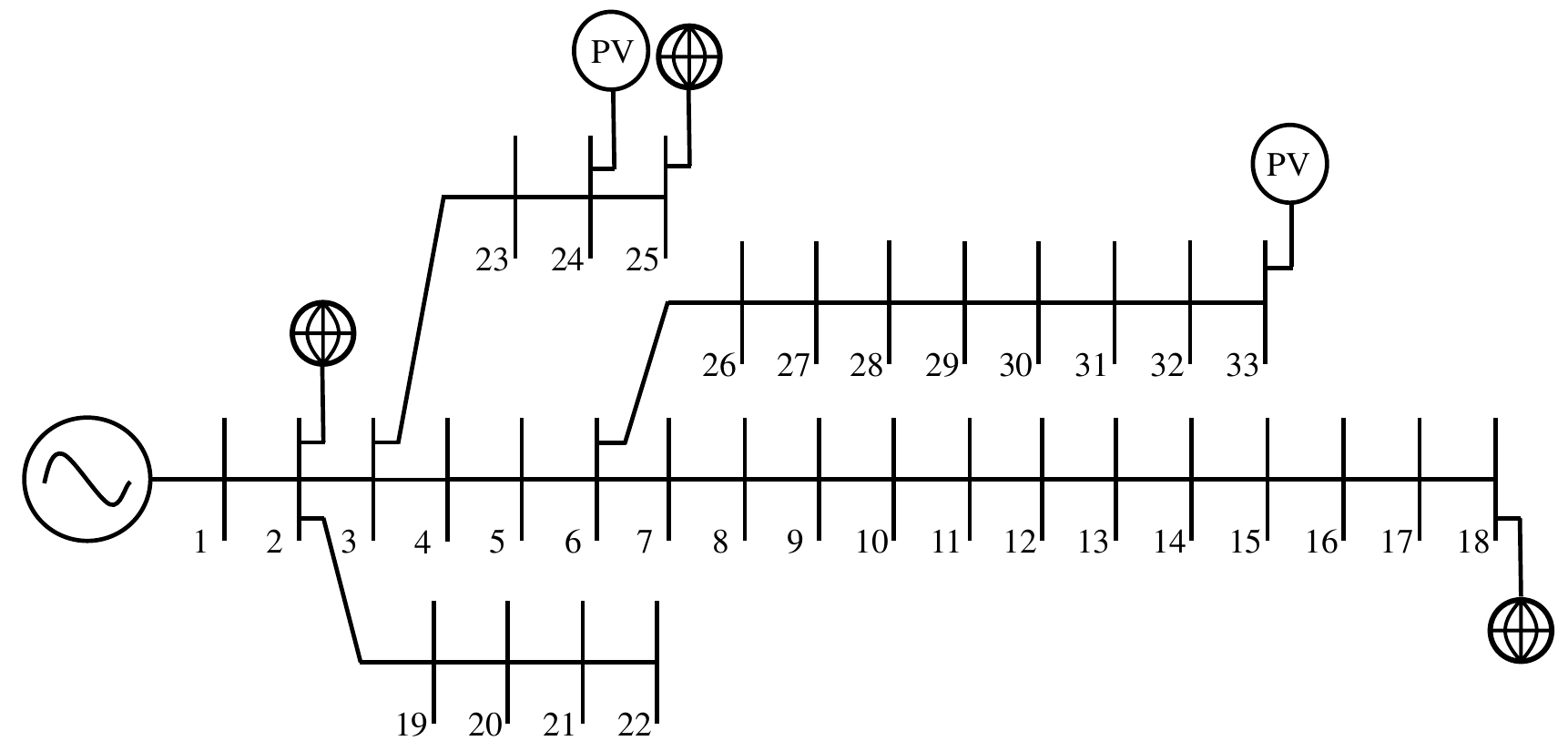}
	\caption{33-bus radial distribution system.}
	\label{fig:33nodes}
\end{figure}

\begin{figure}[!t]
	\centering
	\includegraphics[width=0.45\textwidth]{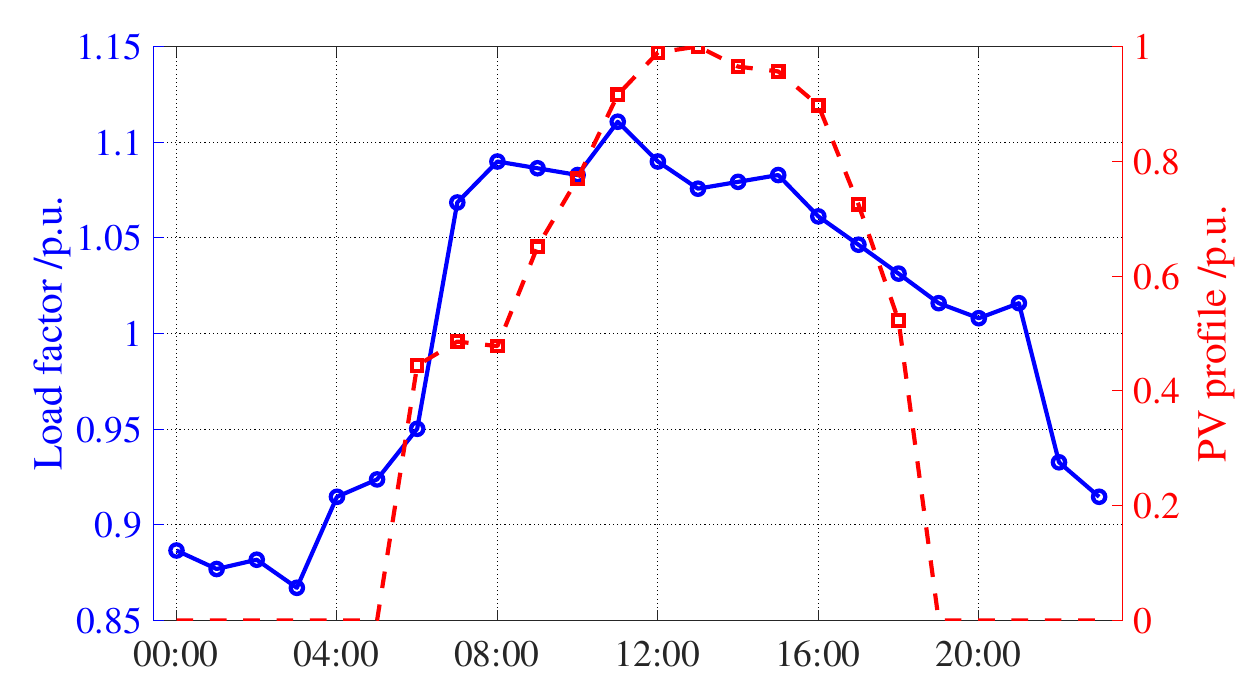}
	\caption{The load factor curve and PV profile.}
	\label{fig:loadfactor}
\end{figure}

In this section, a 33-bus radial distribution system, as depicted in Fig.~\ref{fig:33nodes}, is analyzed to demonstrate the effectiveness of the proposed model~\cite{baran1989network}. The model is implemented in MATLAB and solved using the commercial solver GUROBI. Bus 1 represents the primary substation bus, while nodes 2, 18, and 25 are connected to three IDCs (IDC1, IDC2, and IDC3). All IDCs are operated by a single ISC. Additionally, two photovoltaic (PV) plants are connected to the system at nodes 24 and 33. The per-unit purchased electricity price from the upper grid is 0.075 \$/kWh, and the operation and maintenance (O\&M) cost of the PV plants is 0.047 \$/kWh~\cite{zhang2018distributed}. Twenty-four optimization intervals are considered. The system's load factor and PV load profiles are shown in Fig.~\ref{fig:loadfactor}. 

\begin{table}[!t]
	\renewcommand{\arraystretch}{1.3}
	\caption{ISC-related Paramenters}
	\label{tab:ISC-parameter}
	\centering
	\begin{tabular}{ccc}
		\toprule
		& \thead{10-server 2-workload} & \thead{500-server 6-workload} \\
		\midrule
		\thead{$N^{\mathrm{IT}}_n$} & [10,10,10] & [500,500,500] \\
		\thead{$K_n^C$} & [0.15,0.15,0.15] & [0.15,0.15,0.15] \\
		\thead{$P^{\mathrm{SU}}_{n,s} / P^{\mathrm{SD}}_{n,s}$} & [0,0,0] & [0,0,0] \\
		\thead{$P^{\mathrm{base}}_{n,s}$} & [50,50,50] & [50,50,50] \\
		\thead{$K^{\mathrm{IT}}_{n,s}$} & [1150,1000,1150] & [1150,1000,1150] \\
		\thead{$\Phi_{n,s}$} & [16,16,16] & [32;32;32] \\
		\thead{$\bar{U}_{n}$} & [0.9,0.9,0.9] & [0.9,0.9,0.9] \\
		\thead{$\varphi_t$} & 0.9 & 0.9 \\
		\bottomrule
	\end{tabular}
\end{table}

\subsection{10-server 2-workload Test Case}
\label{sec:10-server-case}
\begin{table}[!t]
	\renewcommand{\arraystretch}{1.3}
	\caption{Workload-related Paramenters for 2-workload Test Case}
	\label{tab:workload-parameter}
	\centering
	\begin{tabular}{ccccc}
		\toprule
		& \thead{$\phi^l$} & \thead{$N^l$} & \thead{$\lambda^{l}_t$} & \thead{$\eta^l$} \\
		\midrule
		\thead{Workload 1} & 4 & 3 & 200 & 0.9 \\
		\thead{Workload 2} & 3 & 4 & 150 & 0.9 \\
		\bottomrule
	\end{tabular}
\end{table}

First, a small-scale scenario where each IDC consists of a limited number of servers (10 servers each) with simple online workload (2 types of workload) is investigated. The purpose of this scenario is to validate the effectiveness of the initial model, the reconstructed model, and the multi-objective optimization framework. The parameters related to the ISC are defined in Table~\ref{tab:ISC-parameter}, while the parameters related to the workload is specified in Table~\ref{tab:workload-parameter}. It's important to note that we do not impose constraints on the number of on-off times during the test, and therefore, $P^{\mathrm{SU}}{n,s} / P^{\mathrm{SD}}{n,s}$ is set to 0. Several cases are compared in this section:

Case 1) The initial model is applied with the DSO's objective $z_2$. In this case, the state of each server is explicitly described by a set of variables, and the ISC is fully controlled by the DSO.

Case 2) The reconstructed model is applied with the DSO's objective $z_2$. In this case, servers are divided into segments based on their operation states, resulting in a reduction of the variable scale.

Case 3) The reconstructed model is applied with the ISC's objective $z_4$. In this case, the priority is given to ensuring the operational efficiency of the ISC. 

Case 4) The reconstructed model is applied using the Nash bargaining framework. In this case, the ISC and DSO act as two bargaining entities to determine the best approach for cooperation.

Table~\ref{tab:results} presents a summary of the overall CPU utilization and DSO cost for the four optimization cases.

\begin{table}[!t]
	\renewcommand{\arraystretch}{1.3}
	\caption{Results of 10-server 2-workload Test Case}
	\label{tab:results}
	\centering
	\begin{tabular}{ccccc}
		\toprule
		& \thead{CPU \\ Utilization (\%)} & \thead{IDC \\ Energy (kWh)} & \thead{DSO \\ Cost (\$)} & \thead{Total \\ Energy (kWh)} \\
		\midrule
		\thead{Case 1} & 73.13 & 657.31 & 5084.85 & 78536.24 \\
		\thead{Case 2} & 73.13 & 657.33 & 5084.86 & 78536.26 \\
		\thead{Case 3} & 84.13 & 676.01 & 5085.59 & 78541.00 \\
		\thead{Case 4} & 79.07 & 658.24 & 5084.98 & 78536.84\\
		\bottomrule
	\end{tabular}
\end{table}

\begin{figure}[!t]
	\centering
	\includegraphics[width=0.45\textwidth]{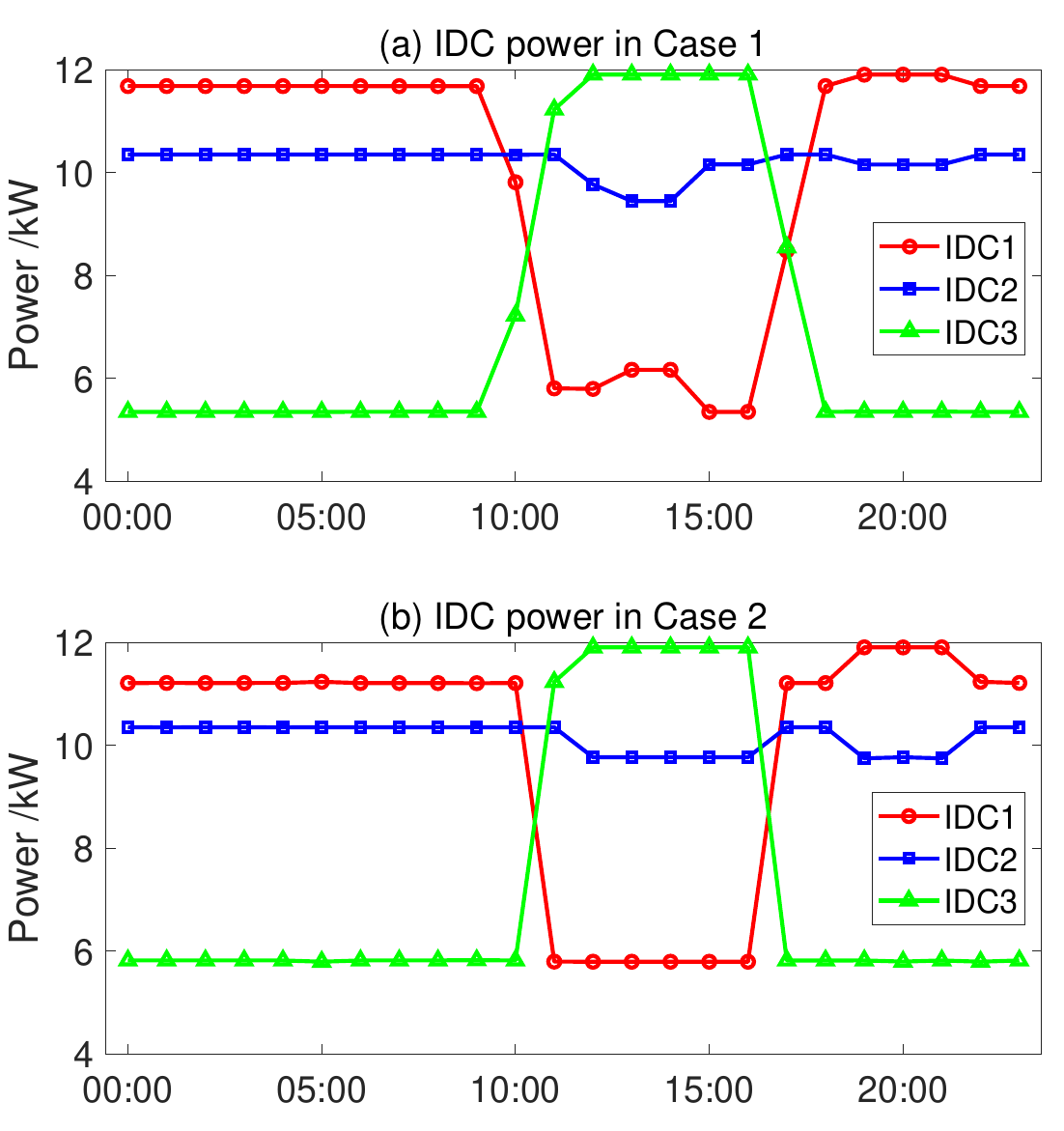}
	\caption{Power consumption of each IDC in Case 1 and Case 2.}
	\label{fig:10-server-idc-power}
\end{figure}

\begin{figure}[!t]
	\centering
	\includegraphics[width=0.45\textwidth]{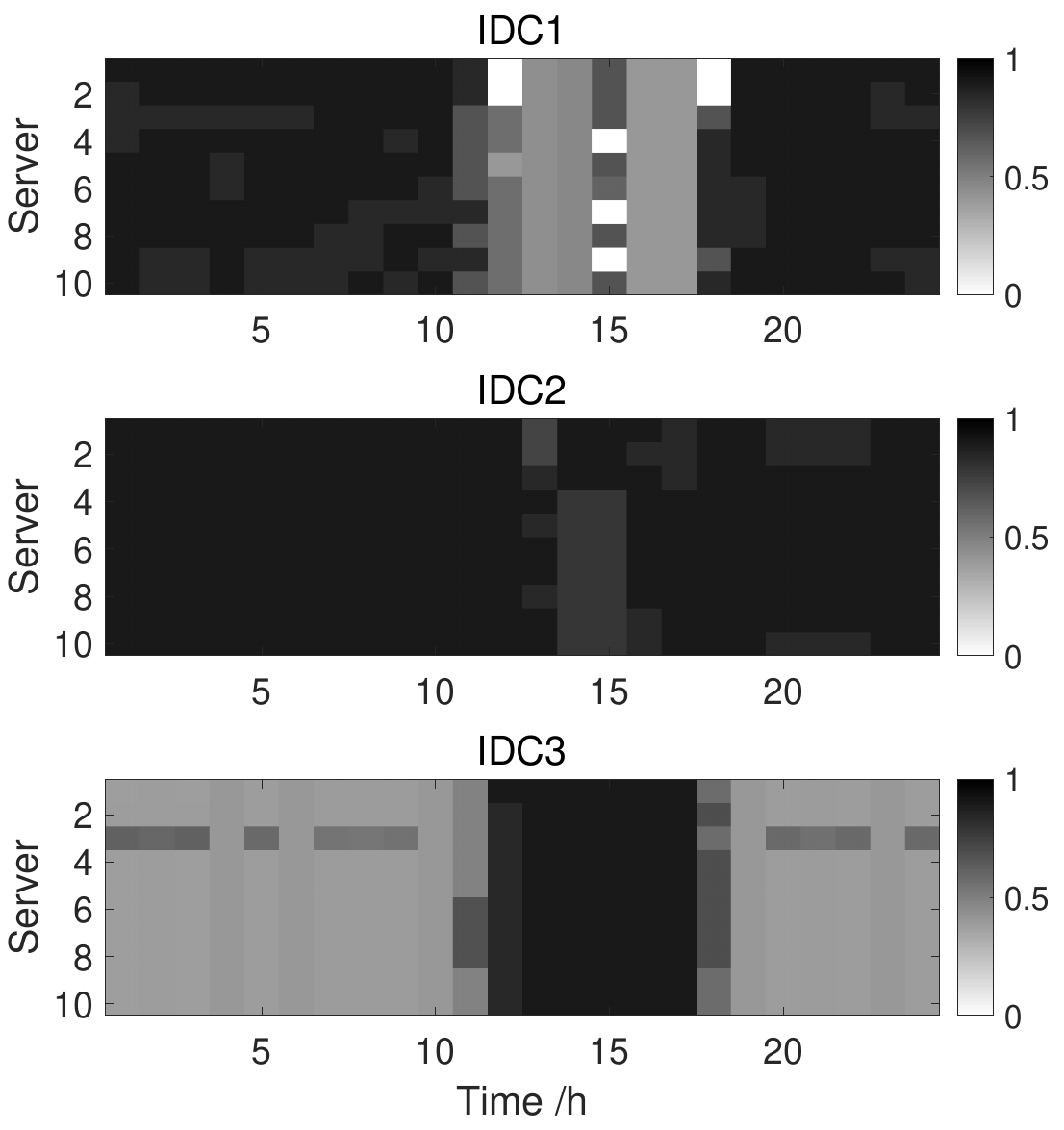}
	\caption{CPU utilization of each time interval in Case 1.}
	\label{fig:10-server-cpu}
\end{figure}

\begin{figure}[!t]
	\centering
	\includegraphics[width=0.45\textwidth]{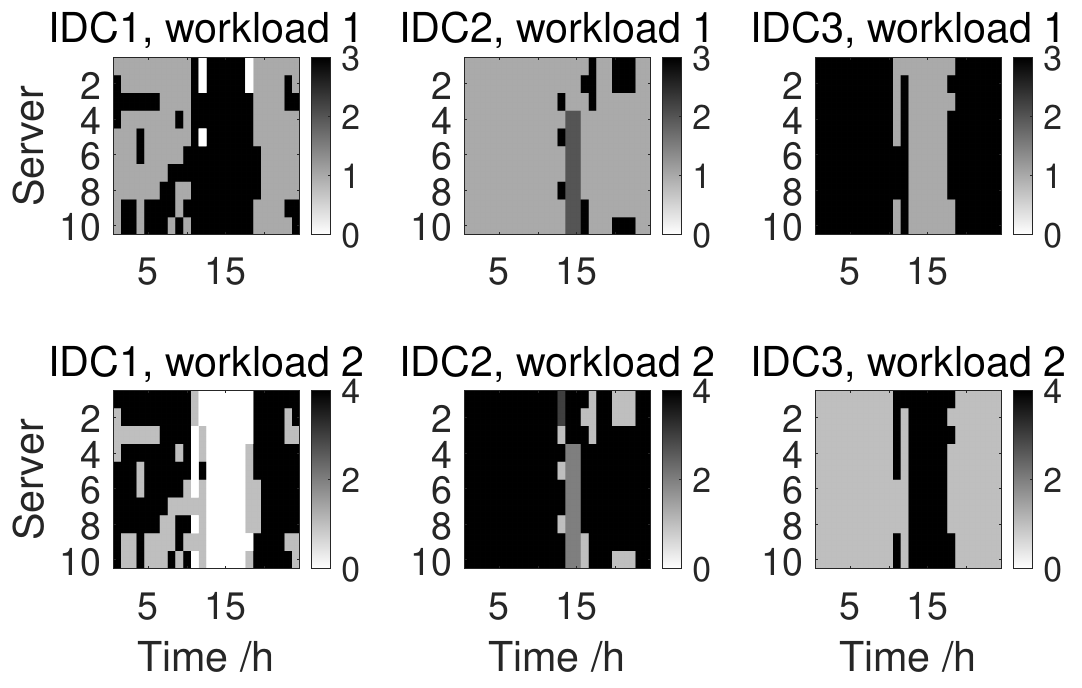}
	\caption{VM deployment of each time interval in Case 1.}
	\label{fig:10-server-workload}
\end{figure}

\textit{1) Effectiveness of Initial Model:} 
In Case 1, the initial model is tested where the ISC is under full control of the DSO. Fig.~\ref{fig:10-server-idc-power} illustrates the optimized IDC power consumption over a 24-hour period. Additionally, Fig.~\ref{fig:10-server-cpu} and Fig.~\ref{fig:10-server-workload} provide visualizations of the server CPU utilization and the VM deployment, respectively. In both figures, the X-axis represents the 24 time periods, while the Y-axis represents the 10 physical servers within each IDC.

In Fig.~\ref{fig:10-server-idc-power}(a), it is observed that during the period from 00:00 to 5:00, when the system load is at its lowest and the PV output is zero, IDC1 and IDC2 consume relatively higher power, while IDC3 consumes relatively lower power. This discrepancy can be attributed to the energy efficiency ratio and the location of each IDC. In Table~\ref{tab:ISC-parameter}, the values of $K^\mathrm{IT}_{n,s}$ are set as [1150, 1000, 1150] for the three IDCs, indicating that servers in IDC2 are more energy-efficient. Consequently, it is a viable strategy to optimize the energy usage mode of the ISC by preferentially distributing the workload to IDC2. However, from the perspective of the DSO, IDC3 is located at the far end of the network, resulting in higher network losses for powering it. Therefore, IDC1 is preferred over IDC3. From Fig.~\ref{fig:10-server-cpu}, it can be observed that initially, IDC2 bears the highest workload, followed by IDC1 and IDC3. After 6:00, as PV output starts to rise along with the system load, the model begins to transfer a portion of the workload from IDC1 to IDC3, as indicated by Fig.~\ref{fig:10-server-idc-power}(a). With the increasing PV output, powering IDC3, which is in close proximity to a PV plant, becomes more cost-effective since PV plants are more economical than purchasing energy from the upper grid. Consequently, around noon, when PV output reaches its peak and surplus power is available, the model transfers part of the workload from IDC2 to IDC1. Additionally, as the system load also reaches its peak, network losses rapidly increase, which also makes processing the workload in IDC2 more expensive than in IDC1. After 17:00, when PV output begins to rapidly decrease, a portion of the workload is transferred back from IDC3 to IDC1. Due to the high system pure load, a small portion of the workload from IDC2 is also transferred to IDC1 due to network losses. Overall, the optimization results of the initial model align with the qualitative analysis.

As discussed in Section~\ref{sec:model-reconstruct}, homogeneity arises when the optimized VM deployment, as determined by the initial model, is replicated across different servers. In Fig.~\ref{fig:10-server-workload}, which illustrates the optimized VM deployment, homogeneity is manifested through repetitive rows, particularly noticeable in IDC2 and IDC3. This observation serves as validation for the foundation of model reconstruction. 

\begin{figure}[!t]
	\centering
	\includegraphics[width=0.45\textwidth]{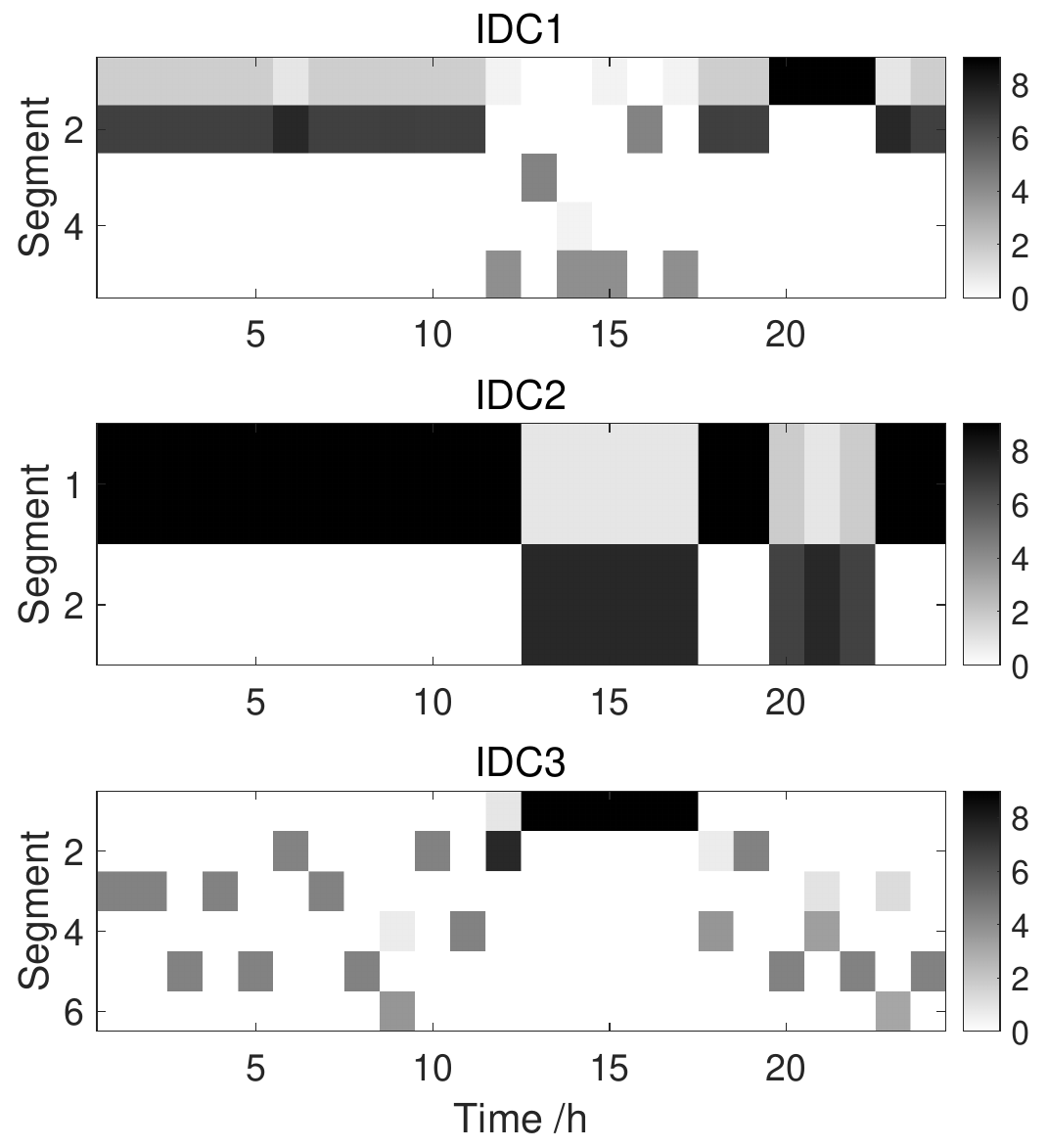}
	\caption{Workload distribution of each time interval in Case 2.}
	\label{fig:10-server-cpu-seg}
\end{figure}

\begin{table}[!t]
	\renewcommand{\arraystretch}{1.3}
	\caption{Generated Deployment Schemes in Case 2}
	\label{tab:10-server-scheme}
	\centering
	\begin{tabular}{ccc}
		\toprule
		\thead{IDC1} & \thead{IDC2} & \thead{IDC3} \\
		\midrule
		1,4 & 1,4 & 1,4\\
		3,1 & 3,1 & 3,1 \\
		2,2 & & 2,2 \\
		0,4 & & 3,0 \\
		3,0 & & 2,1 \\
		& & 2,0 \\
		\bottomrule
	\end{tabular}
\end{table}

\textit{2) Exactness of Model Reconstruction:} 
In Case 2, the reconstructed model is tested under the same conditions as in Case 1. Fig.~\ref{fig:10-server-idc-power}(b) illustrates the power consumption of each IDC in Case 2. Comparing it with Fig.~\ref{fig:10-server-idc-power}(a), Fig.~\ref{fig:10-server-idc-power}(b) demonstrates a similar workload dispatch strategy, but with faster and smoother workload transfers among IDCs. From Table~\ref{tab:results}, it is evident that the DSO costs of Case 1 and Case 2 are approximately the same. The difference between the two optimal objective values is $1.97 \times 10^{-6}$, which confirms the accuracy of the reconstructed model.

Unlike the initial model that considers servers individually, the reconstructed model divides the server cluster into segments. Each segment corresponds to a deployment scheme generated by the algorithm proposed in Section~\ref{sec:algorithm}. For each IDC, there are 14 deployment schemes that meet the requirements, while the reconstructed model generates 5, 2, and 6 schemes for IDC1, IDC2, and IDC3, respectively. These generated schemes are listed in Table~\ref{tab:10-server-scheme}. Each scheme is represented by an ordered number pair, indicating the number of deployed VMs for two types of workload on a server. This result supports the assumption of the reconstructed model that only a small number of schemes need to be considered. The optimized workload distribution is depicted in Fig.~\ref{fig:10-server-cpu-seg}. The color representation in the figure reflects the aggregated CPU utilization of a segment (CPU utilization multiplied by the batch size), while the X-axis corresponds to the 24 time periods and the Y-axis represents the different segments within each IDC. It can be observed that Fig.~\ref{fig:10-server-cpu-seg} exhibits greater clarity compared to Fig.~\ref{fig:10-server-cpu} due to the absence of repetitive rows. This indicates that the model reconstruction reduces the redundancy of variables and effectively decreases the scale of the problem.

\begin{figure}[!t]
	\centering
	\includegraphics[width=0.45\textwidth]{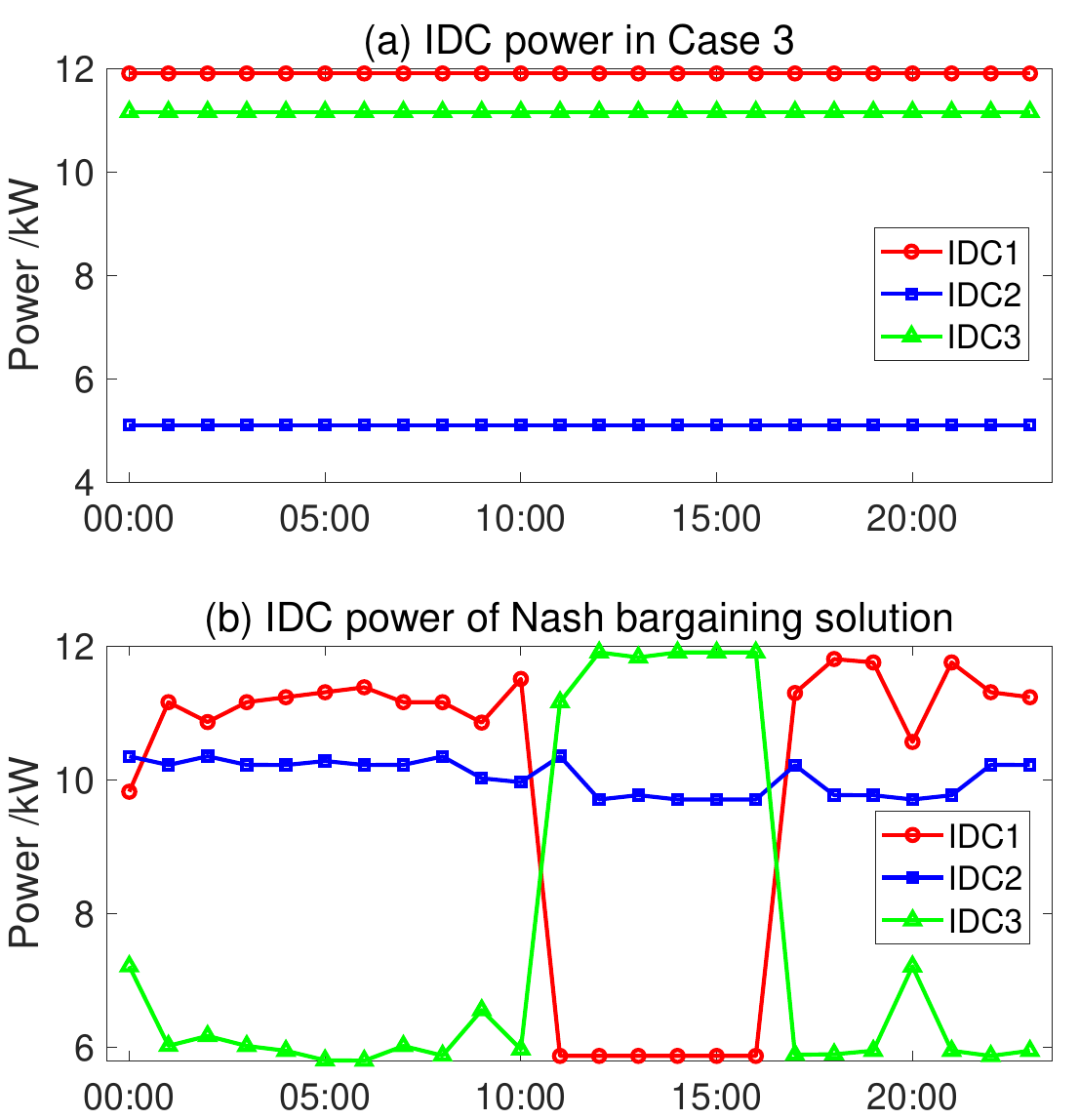}
	\caption{Power consumption of each IDC in Case 3 and Case 4.}
	\label{fig:10-server-idc-power-nash}
\end{figure}

\begin{figure}[!t]
	\centering
	\includegraphics[width=0.45\textwidth]{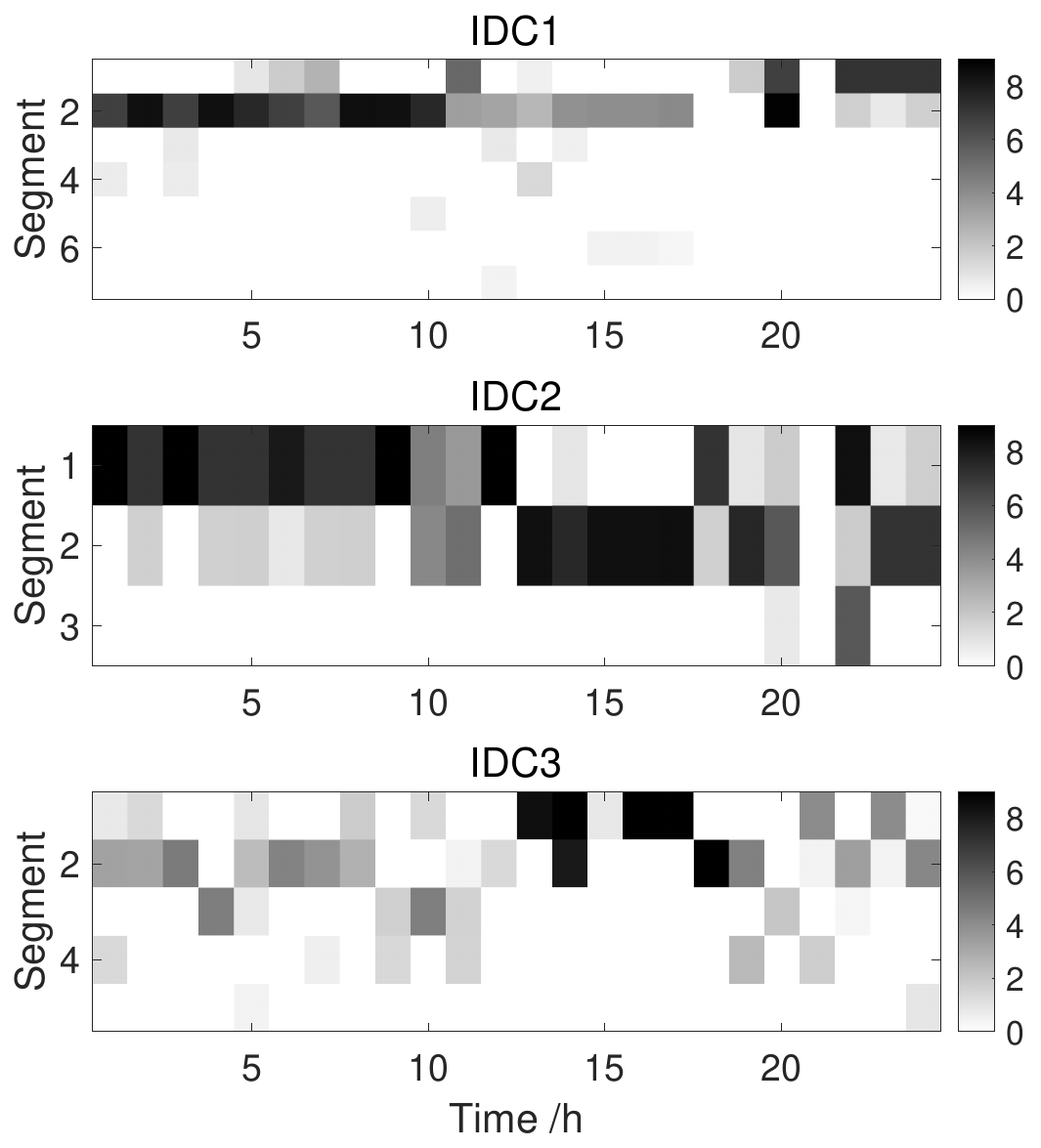}
	\caption{Workload distribution of Nash bargaining solution}
	\label{fig:10-server-cpu-nash}
\end{figure}

\begin{figure}[!t]
	\centering
	\includegraphics[width=0.45\textwidth]{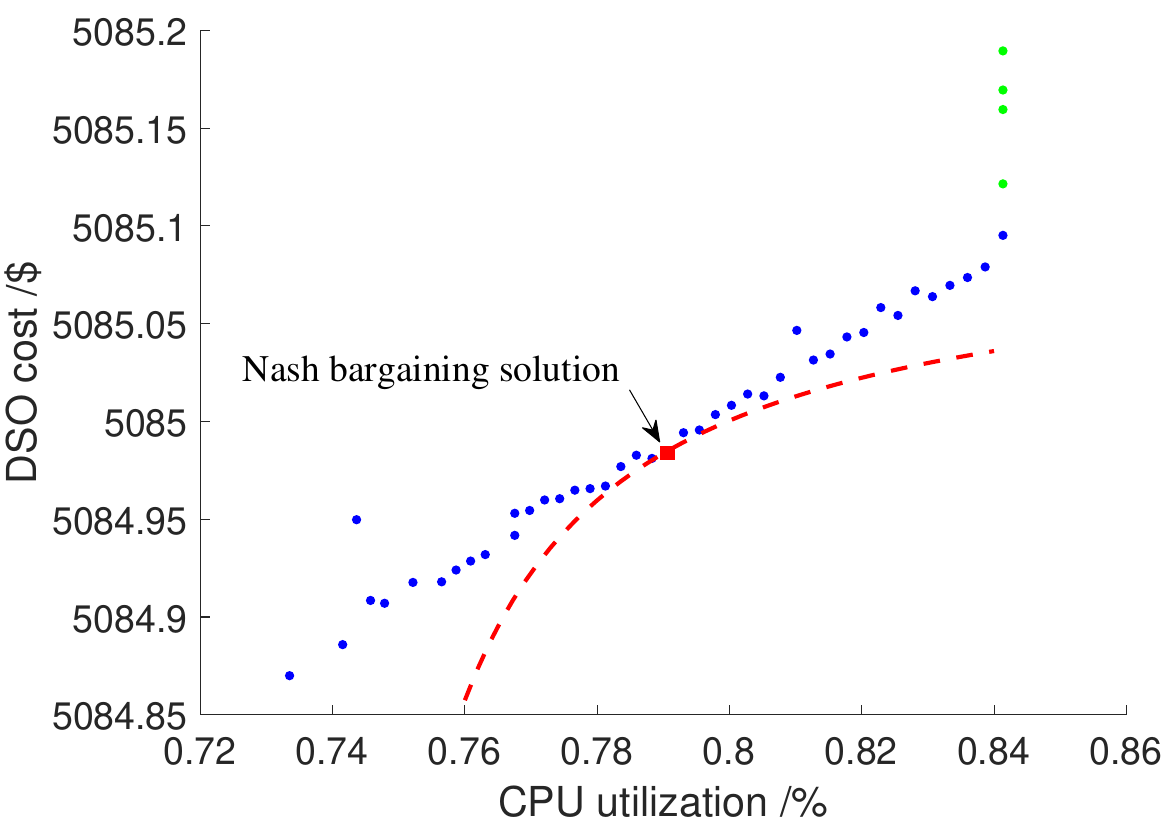}
	\caption{Pareto front and Nash bargaining solution.}
	\label{fig:pareto}
\end{figure}

\textit{3) Evaluation of Nash Bargaining Framework:} 
Case 2 and Case 3 are implemented using the objective functions of DSO and ISC, respectively, within the same model and parameter settings. Comparing the optimization results in Table~\ref{tab:results}, a prominent conflict between the two objectives of DSO and ISC becomes apparent. Case 3 indicates that the maximum CPU utilization of the ISC is 84.13\%, which is 11.00\% higher than that in Case 2, where the ISC's objective is not considered. This in turn leads to an increase of 0.73 \$ in DSO generation costs, which accounts for 0.014\% of the total cost in Case 2. Fig.~\ref{fig:10-server-idc-power-nash}(a) illustrates the optimized IDC power consumption in Case 3. The workload dispatch strategy in this case significantly differs from that in Case 2. The ISC distributes the workload among different IDCs with the goal of improving CPU utilization, regardless of system load pressure and energy efficiency, which is less cost-effective for the DSO. Additionally, Case 4 employs the Nash bargaining framework to provide further insight. Fig.~\ref{fig:10-server-idc-power-nash}(b) displays the IDC power consumption for the Nash bargaining solution, while Fig.~\ref{fig:10-server-cpu-nash} depicts the corresponding workload distribution. The results demonstrate that, unlike Case 3, the Nash bargaining solution adopts a strategy similar to Case 2 while also considering the interests of the ISC. As shown in Fig.~\ref{fig:10-server-idc-power-nash}(b), based on the scheduling strategy in Case 2, the Nash bargaining solution transfers a portion of the workload from IDC1 to IDC3 during idle hours of the system. Intuitively, evenly distributing the workload among all IDCs is more conducive to improving average CPU utilization. Since IDC3 is less heavily loaded during these hours, this strategy helps improve overall CPU utilization, which aligns with the results presented in Table~\ref{tab:results}.

Fig.~\ref{fig:pareto} displays the Pareto front $\mathcal{P}$ and the Nash bargaining solution obtained from optimization problem~(\ref{model:nash}). The blue dots represent solutions on the Pareto front, while the green dots represent solutions that are not Pareto-efficient. As indicated by the figure, the Pareto-efficient DSO cost when CPU utilization is maximized is 5085.10 \$, indicating that the solution given by Case 3 is not Pareto-efficient. This observation validates that the ISC can dispatch workload within a certain range without incurring additional costs. As indicated by the green dots in Fig.~\ref{fig:pareto}, in the first stage when the ISC starts participating in cooperative optimization, the DSO cost is reduced from 5085.59 \$ to 5085.10 \$, representing a decrease of 0.010\%, while maintaining the ISC's CPU utilization at 84.13\%. Subsequently, in the second stage, both objectives start to decrease simultaneously, as depicted by the blue dots. In the Nash bargaining solution, the CPU utilization is maintained at 79.07\%, while the DSO cost is further reduced from 5085.10 \$ to 5084.98 \$, a decrease of 0.002\%. Throughout this stage, the ISC optimizes its power consumption, saving 17.77 kWh of electricity, which accounts for 2.63\% of its overall energy consumption. Additionally, it optimizes the system power flow to enhance the absorption of PV, resulting in a 0.012\% reduction in DSO costs. However, it is not in the best interest of the entire system to further decrease the DSO cost by 0.002\%, as it would lead to a 5.94\% decrease in CPU utilization, which aligns with common intuition.

\begin{table}[!t]
	\renewcommand{\arraystretch}{1.3}
	\caption{Workload-related Paramenters for 500-server Test Case}
	\label{tab:6-workload-parameter}
	\centering
	\begin{tabular}{ccccc}
		\toprule
		& \thead{$\phi^l$} & \thead{$N^l$} & \thead{$\lambda^{l}_t$} & \thead{$\eta^l$} \\
		\midrule
		\thead{Workload 1} & 8 & 3 & 3600 & 0.9 \\
		\thead{Workload 2} & 4 & 1 & 3000 & 0.9 \\
		\thead{Workload 3} & 4 & 3 & 5000 & 0.9 \\
		\thead{Workload 4} & 8 & 3 & 3600 & 0.9 \\
		\thead{Workload 5} & 4 & 3 & 4000 & 0.9 \\
		\thead{Workload 6} & 8 & 3 & 3000 & 0.9 \\
		\bottomrule
	\end{tabular}
\end{table}

\subsection{500-server 6-workload Test Case}
\begin{figure}[!t]
	\centering
	\includegraphics[width=0.45\textwidth]{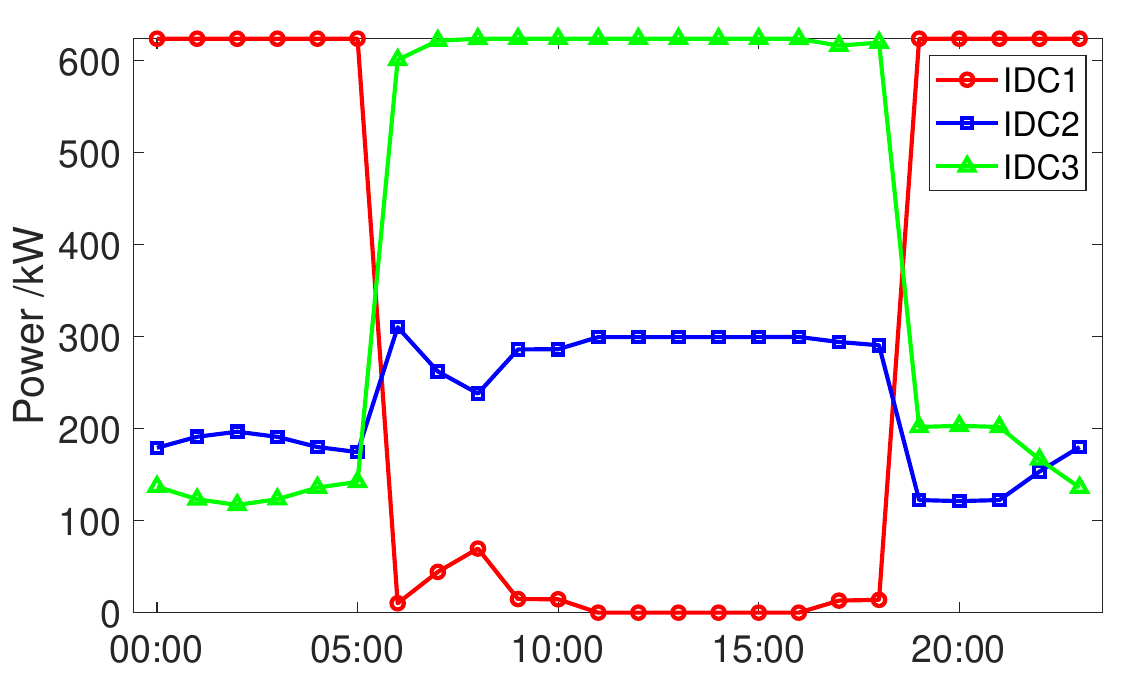}
	\caption{Power consumption of each IDC in 500-server test case.}
	\label{500-server-idc-power}
\end{figure}

\begin{figure}[!t]
	\centering
	\includegraphics[width=0.45\textwidth]{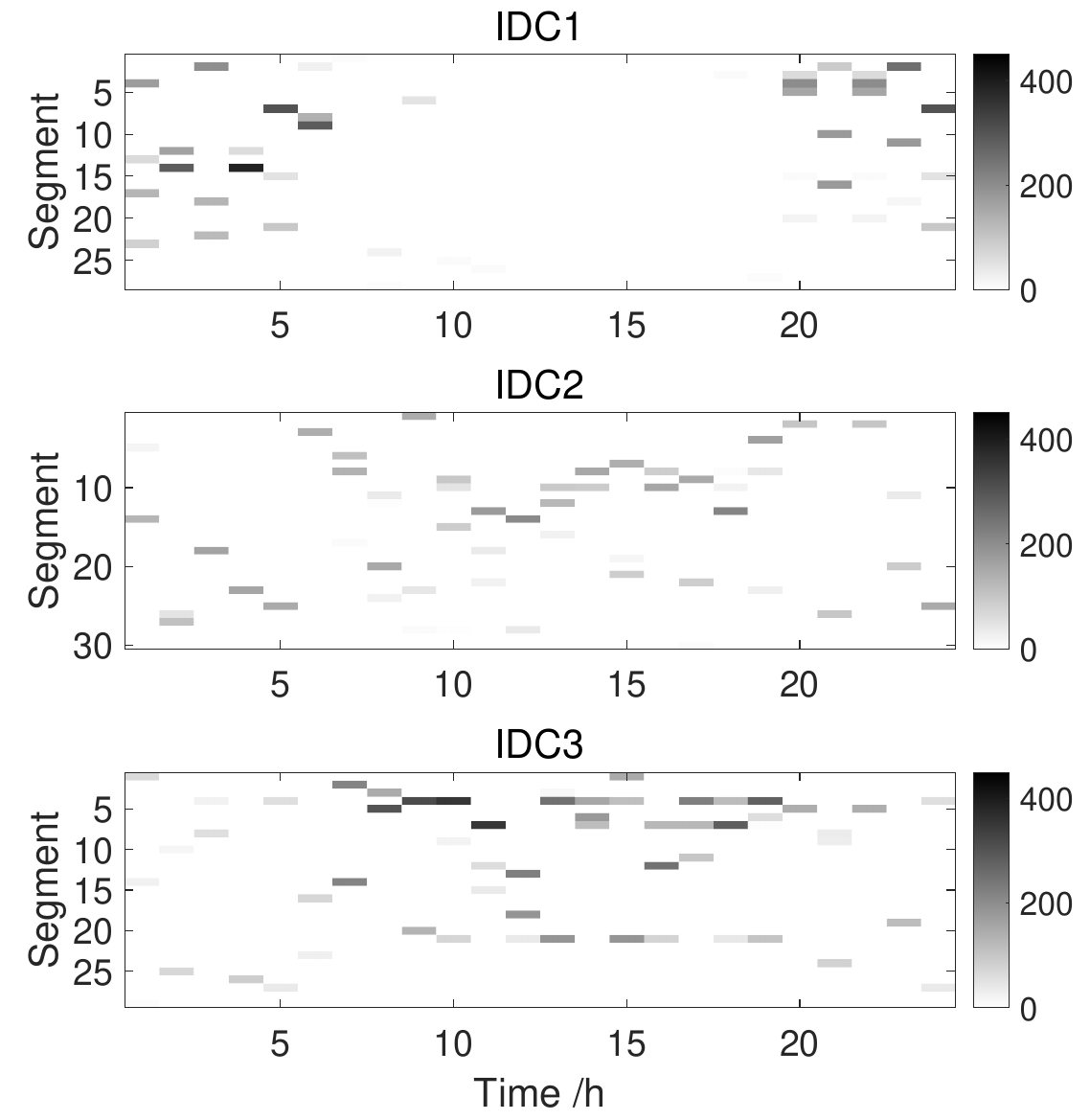}
	\caption{Workload distribution in 500-server test case.}
	\label{500-server-cpu}
\end{figure}

A larger test case consisting of 500 servers and 6 types of workload are analyzed to further demonstrate the effectiveness of the proposed model. This test case specifically aims to evaluate the solution algorithm of the reconstructed model. Therefore, the reconstructed model with the Nash bargaining solution is adopted. The ISC-related parameters are specified in Table~\ref{tab:ISC-parameter}, while the workload-related parameters are defined in Table~\ref{tab:6-workload-parameter}.

Fig.~\ref{500-server-idc-power} and Fig.~\ref{500-server-cpu} illustrate the results of the test case. In this particular test case, an overall CPU utilization of 83.13\% is achieved for the ISC, while the total cost for the DSO amounts to 5229.62 \$. As shown in Fig.~\ref{500-server-idc-power}, the scheduling strategy employed by the model remains similar to Case 4 in Section~\ref{sec:10-server-case}, where the Nash bargaining framework was also utilized.

Regarding the deployment schemes, the algorithm generated 28, 30, and 29 schemes for each IDC, respectively. Additionally, the total number of feasible deployment schemes for each IDC reaches 364. This outcome indicates that the proposed algorithm can effectively reduce the variable scale by more than one-tenth.

\section{Conclusion}
\label{sec:conclusion}
This paper focuses on utilizing the ISC as a specialized flexible resource to actively participate in collaborative optimization for grid power flow. In this study, a computation-power coupling model is developed, incorporating the mechanisms of VM deployment and workload allocation. Furthermore, an ISC-DSO collaborative optimization model is constructed, accompanied by a corresponding solution algorithm to address computational challenges. To address the conflicting objectives between the ISC and DSO, a multi-objective optimization framework is designed based on the characteristics of the Nash bargaining solution. This framework effectively avoids subjective randomness in the selection of optimal points. A typical 33-node distribution network is adopted to conduct the case study. In the small-scale case, the results demonstrate that by reducing the CPU utilization of the ISC by 5.06\%, a reduction of 0.012\% in the total DSO cost can be achieved. Furthermore, the larger case corroborates these findings, further validating the accuracy of the model reconstruction and the effectiveness of the solution algorithm.

% Can use something like this to put references on a page
% by themselves when using endfloat and the captionsoff option.
\ifCLASSOPTIONcaptionsoff
\newpage
\fi

% trigger a \newpage just before the given reference
% number - used to balance the columns on the last page
% adjust value as needed - may need to be readjusted if
% the document is modified later
%\IEEEtriggeratref{8}
% The "triggered" command can be changed if desired:
%\IEEEtriggercmd{\enlargethispage{-5in}}

% references section

% can use a bibliography generated by BibTeX as a .bbl file
% BibTeX documentation can be easily obtained at:
% http://mirror.ctan.org/biblio/bibtex/contrib/doc/
% The IEEEtran BibTeX style support page is at:
% http://www.michaelshell.org/tex/ieeetran/bibtex/
%\bibliographystyle{IEEEtran}
% argument is your BibTeX string definitions and bibliography database(s)
%\bibliography{IEEEabrv,../bib/paper}
%
% <OR> manually copy in the resultant .bbl file
% set second argument of \begin to the number of references
% (used to reserve space for the reference number labels box)
\bibliographystyle{IEEEtran}
\bibliography{ref}

\begin{IEEEbiography}[{\includegraphics[width=1in,height=1.25in,clip,keepaspectratio]{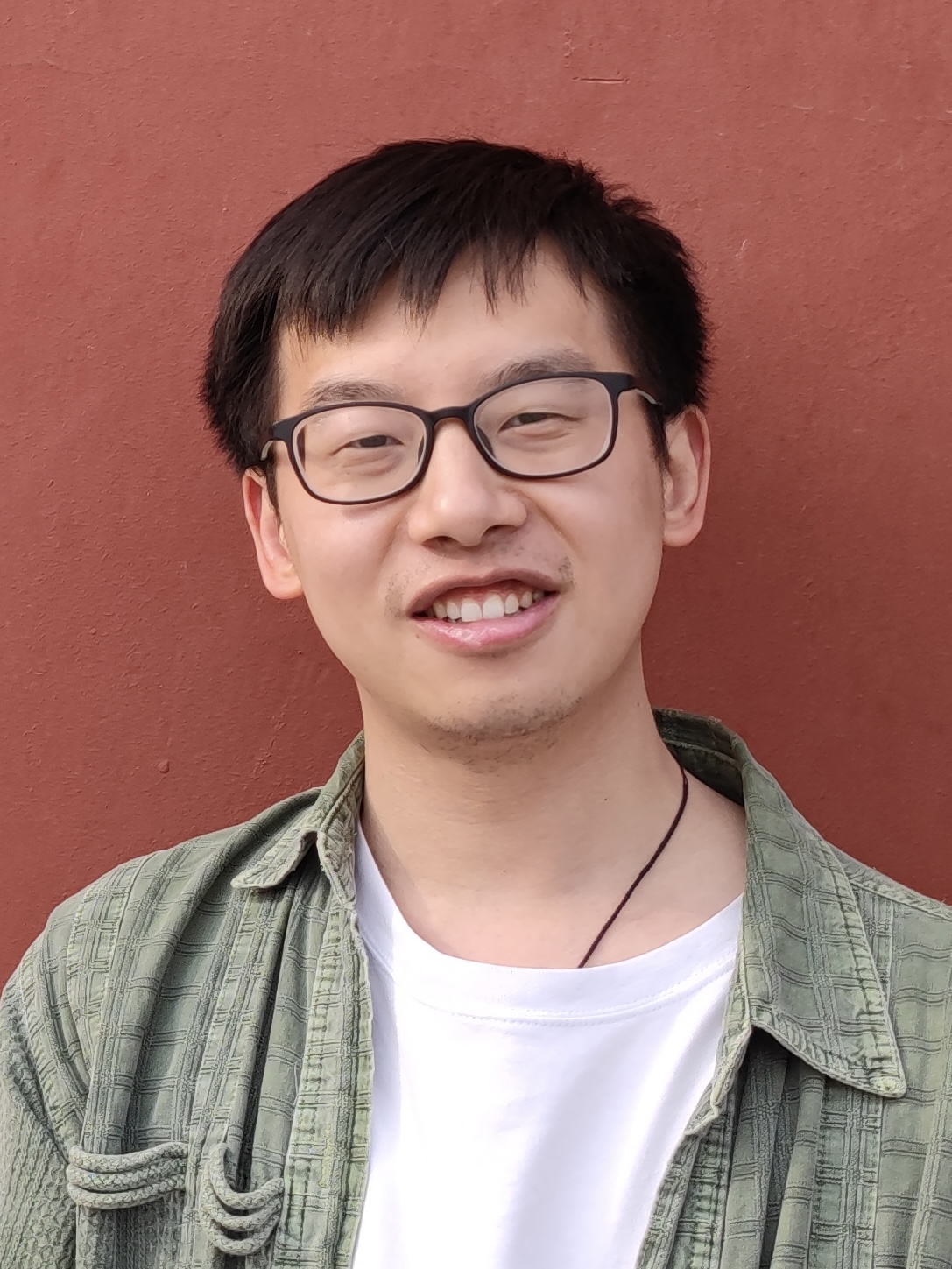}}]{Chuyi Li} received the B.S. degree in electrical engineering from Huazhong University of Science and Technology, Wuhan, China, in 2021.
	He is currently pursuing the Ph.D. degree in electrical engineering with Tsinghua University, Beijing, China. His research interests include load data analysis and virtual power plant flexible resource modeling. 
\end{IEEEbiography}

\begin{IEEEbiography}[{\includegraphics[width=1in,height=1.25in,clip,keepaspectratio]{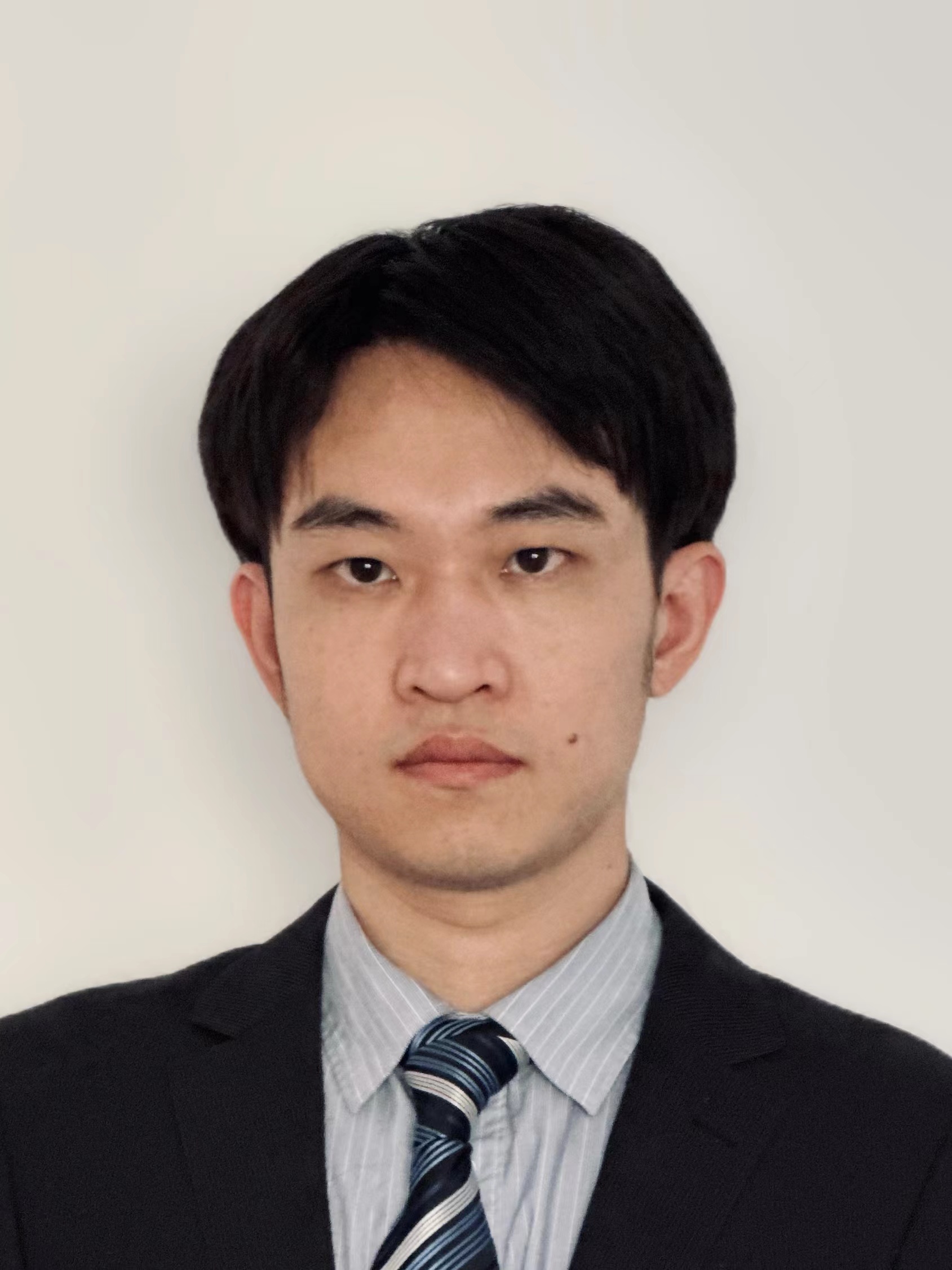}}]{Kedi Zheng} (Member, IEEE) received the B.S. and Ph.D. degrees in electrical engineering from Tsinghua University, Beijing, China, in 2017 and 2022, respectively.
	He is currently a Postdoctoral Researcher with Tsinghua University. He is also a Visiting Research Associate with The University of Hong Kong. His research interests include data analytics in power systems and electricity markets. His research interests include application of big data analytics for electricity market.
\end{IEEEbiography}

\begin{IEEEbiography}[{\includegraphics[width=1in,height=1.25in,clip,keepaspectratio]{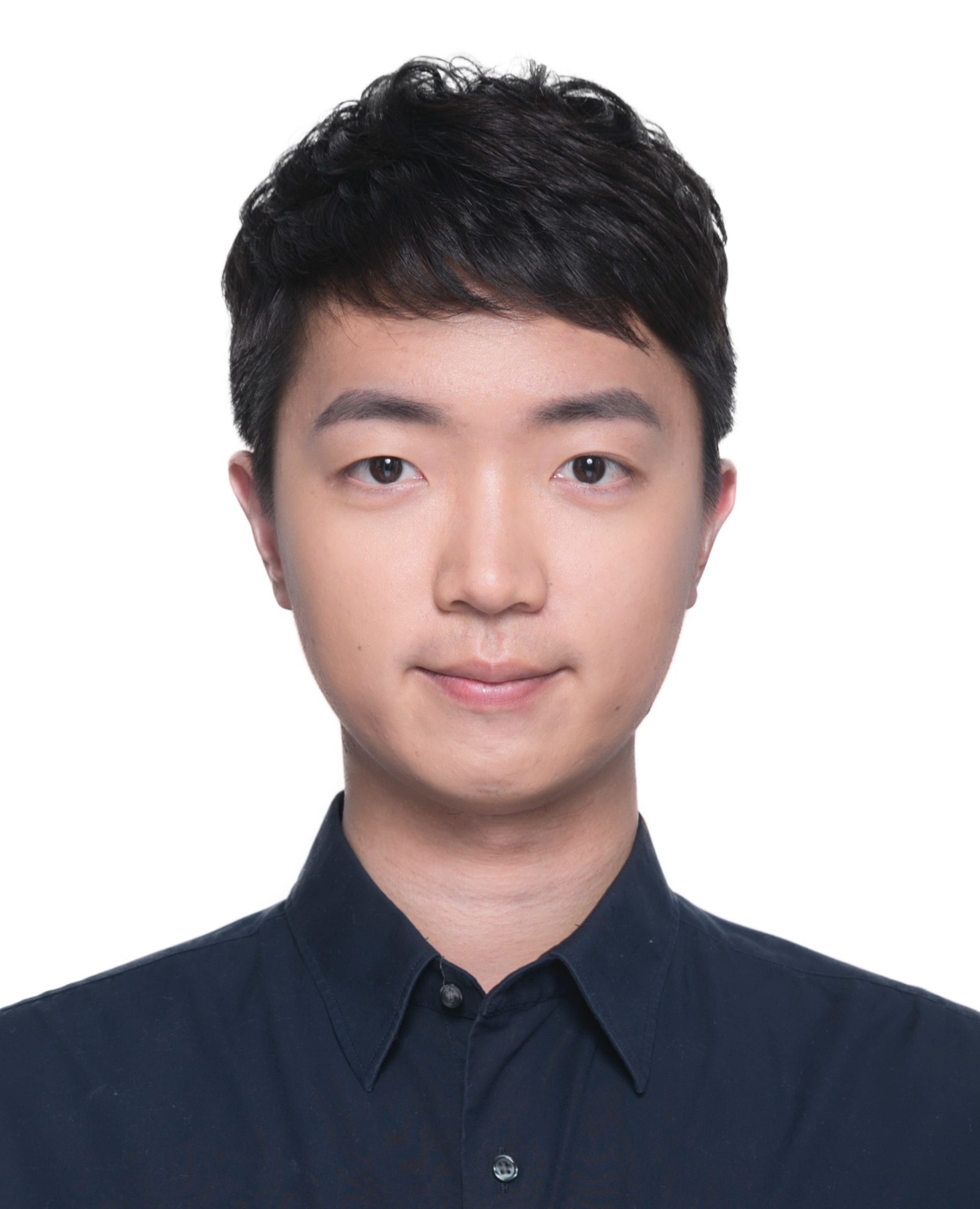}}]{Hongye Guo} (Member, IEEE) received the B.S. and Ph.D. degrees in electrical engineering from Tsinghua University, Beijing, China, in 2015 and 2020, respectively. 
	He is currently a Postdoctoral Research Fellow with Tsinghua University. His research interests include electricity markets, game theory, energy economics, and machine learning.
\end{IEEEbiography}

\begin{IEEEbiography}[{\includegraphics[width=1in,height=1.25in,clip,keepaspectratio]{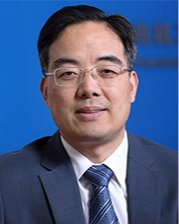}}]{Chongqing Kang} (Fellow, IEEE) received the Ph.D. degree from the Department of Electrical Engineering, Tsinghua University, Beijing, China, in 1997. 
	He is currently a Professor with Tsinghua University. His research interests include power system planning, power system operation, renewable energy, low-carbon electricity technology, and load forecasting.
\end{IEEEbiography}

\begin{IEEEbiography}[{\includegraphics[width=1in,height=1.25in,clip,keepaspectratio]{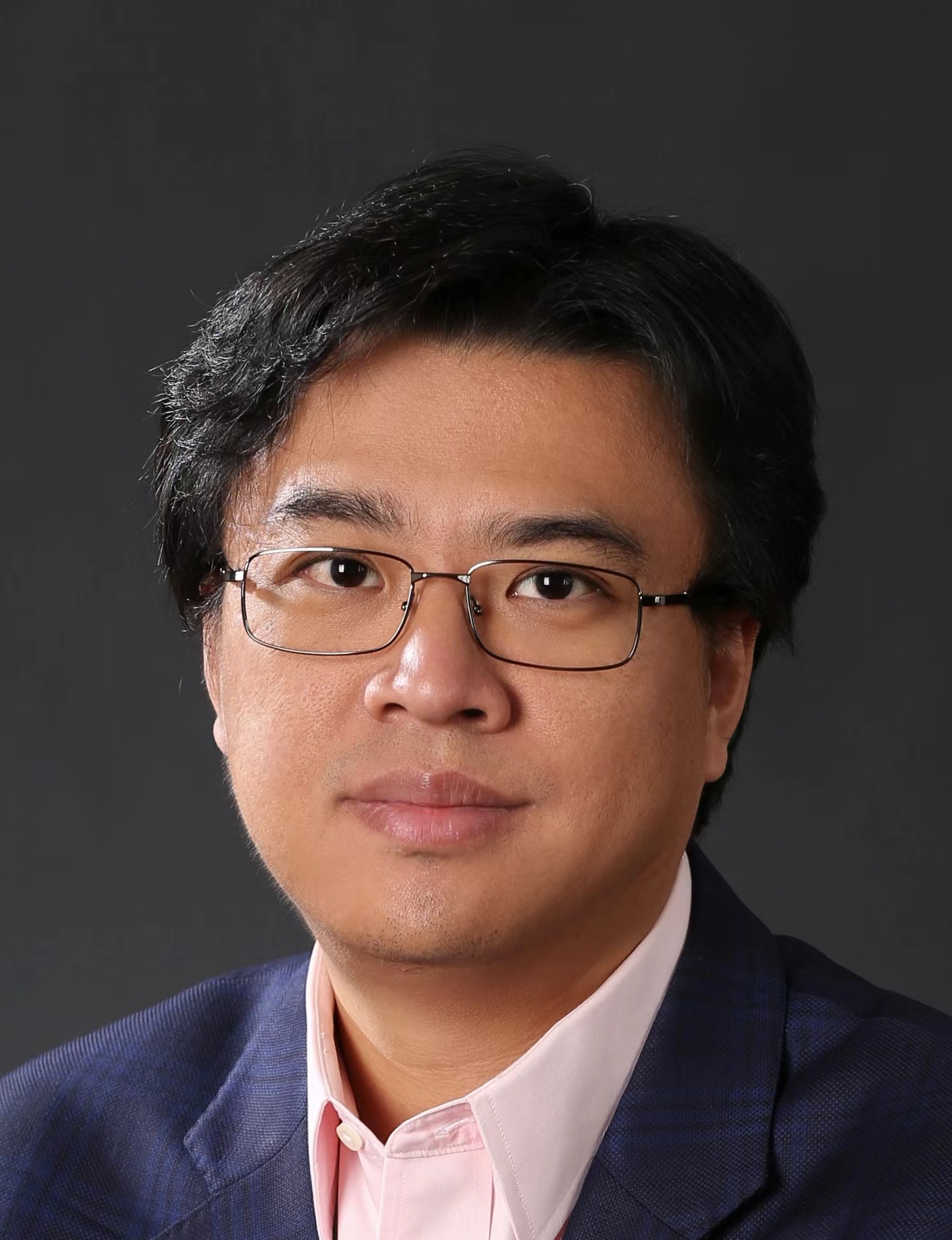}}]{Qixin Chen} (Senior Member, IEEE) received the Ph.D. degree from the Department of Electrical Engineering, Tsinghua University, Beijing, China, in 2010, where he is currently a Tenured Professor. 
	His research interests include electricity markets, power system economics and optimization, low-carbon electricity, and data analytics in power systems.
\end{IEEEbiography}

\vfill

% Can be used to pull up biographies so that the bottom of the last one
% is flush with the other column.
%\enlargethispage{-5in}

% that's all folks
\end{document}